\documentclass{article}
\usepackage{natbib,epsfig,multirow}

\usepackage{graphicx,natbib}
\usepackage{mathtools}
\usepackage{soul}

\usepackage{amsmath,placeins}
\usepackage{bm}
\usepackage{dsfont}
\usepackage{tikz}
\usetikzlibrary{snakes}
\usetikzlibrary{arrows}
\usetikzlibrary{shapes}
\usetikzlibrary{backgrounds}
\usepackage{setspace}
\newcommand\mat[1]{\mathds{#1}}
\newcommand{\bs}{\boldsymbol}
\newcommand{\bTheta}{{\bs\Theta}}

\newcommand{\mx}{\mat{X}}
\newcommand{\my}{\mat{Y}}
\newcommand{\mz}{\mat{Z}}
\newcommand{\x}{\textbf{\textit{x}}}

\newcommand{\sumg}{\sum_{g=1}^G}
\newcommand{\sumjone}{\sum_{j=1}^{n_1}}

\newcommand{\sumj}{\sum_{j=1}^{n_i}}

\newcommand{\sumi}{\sum_{i=1}^2}

\newcommand{\prodg}{\prod_{g=1}^G}
\newcommand{\prodjone}{\prod_{j=1}^{n_1}}
\newcommand{\prodjtwo}{\prod_{j=1}^{n_2}}
\newcommand{\prodim}{\prod_{i=1}^m}
\newcommand{\prodj}{\prod_{j=1}^{n_i}}
\newcommand{\prodi}{\prod_{i=1}^2}

\newcommand{\zigone}{z_{jg}^{(1)}}

\newcommand{\zighat}{\hat{z}_{jg}^{(i)}}

\newcommand{\zighattwo}{\hat{z}_{jg}^{(2)}}
\newcommand{\kout}{^{-(k)}}

\newcommand{\covar}{\bs{\Sigma}}
\newcommand{\ptp}{^{(t+1)}}
\newcommand{\pt}{^{(t)}}

\newcommand{\lik}[2]{\mathcal L_{\text{#1}}(\bTheta \mid {#2})}
\newcommand{\likhat}[2]{\mathcal L_{\text{#1}}(\hat\bTheta \mid {#2})}
\newcommand{\mixx}[1]{\pi_{g} \phi(\bs x_{{#1}}\mid {\bs\theta}_{g})}
\newcommand{\mixt}[2]{\pi_{g}^{({#2})}\phi(\bs x_{{#1}}\mid {\bs\theta}^{({#2})}_{g})}
\newcommand{\mixhat}[1]{\hat{\pi}_{g} \phi(\bs x_{{#1}}\mid \hat{\bs\theta}_{g})}
\newcommand{\wLc}{\mathcal L_\text{w}(\bTheta\mid D_\text{c})}

\newcommand{\wL}{\mathcal L_\text{w} (\bTheta \mid D_\text{o})}
\newcommand{\wl}{\ell_\text{w}(\bTheta \mid D_\text{o})}

\newcommand{\hz}{\hat{z}_{jg}^{(i)}}

\newcommand{\Lclust}{\lik{\text{Clst}}{D_\text{u}}}
\newcommand{\Lda}{\lik{\text{DA}}{D_\text{L}}}
\newcommand{\Lsemi}{\lik{\text{semi}}{D_\text{o}}}
\newcommand{\Lclusthat}{\likhat{\text{Clst}}{D_\text{u}}}
\newcommand{\Ldahat}{\likhat{DA}{D_\text{L}}}
\newcommand{\Lsemihat}{\likhat{\text{semi}}{D_\text{o}}}

\newcommand{\fonehat}{f_1(\x \mid \hat\bTheta)}
\newcommand{\ftwohat}{f_2(\x \mid \hat\bTheta)}
\newcommand{\ftheta}[1]{f_{#1}(\x \mid \bTheta)}

\newcommand{\fcompj}[1]{f_1(\bs x_{{#1}j}, \bs z_{{#1}j} \mid \bTheta_1)}

\newcommand{\fonexzi}{f_1(\mx_{i}, \mz_i \mid \bTheta_1)}

\newcommand{\fhatc}[1]{f_{#1}(\x \mid \hat\bTheta_{\ac})}
\newcommand{\fhat}[1]{\text{FSC}_{#1}}

\newcommand{\rw}{\omega}

\newcommand{\ac}{{\rw_{c}}}
\newcommand{\ai}{{\rw_{i}}}
\newcommand{\aone}{{\rw_{1}}}
\newcommand{\atwo}{{\rw_{2}}}

\newcommand{\aopt}{\abold^{\text{opt}}}
\newcommand{\abold}{\boldsymbol{\rw}}

\newcommand{\awhatvec}{\boldsymbol\rw_{A}}
\newcommand{\awhat}{\rw_{a}}
\newcommand{\am}[1]{{\rw_{#1}}}

\newcommand{\aklopt}[1]{\hat\rw_{\text{KL}_{#1}}}

\newcommand{\fda}{\text{FSC}_{\text{DA}}}
\newcommand{\fclass}{\text{FSC}_{\text{class}}}
\newcommand{\fclust}{\text{FSC}_{\text{clust}}}

\newcommand{\frhat}{\text{FSC}_{\awhat}}
\newcommand{\fari}{\text{FSC}_{\text{ARI}}}

\newcommand{\dataD}{ D_{\Delta rp}}
\newcommand{\avgari}{\overline{\text{ARI}}(\awhat, D_{\cdot p})}
\newcommand{\avgariD}{\overline{\text{ARI}}(\awhat, D_{\Delta \cdot p})}
\newcommand{\ariopt}{\hat\rw_{\text{ARI}_p}}

\begin{document}

\title{Mixture Model Averaging for Clustering}
\author{Irene Vrbik\thanks{Department of Mathematics \& Statistics, McGill University, 805 Sherbrooke Street West, Montreal, Quebec H3A 0B9. Email {\tt irene.vrbik@mcgill.ca}} \ and Paul~D.~McNicholas\thanks{Department of Mathematics \& Statistics, McMaster University, Hamilton, Ontario, Canada,  L8S~4L8. Tel.: 905-525-9140, ext.\ 23419. Email {\tt mcnicholas@math.mcmaster.ca}}}
\date{}
\maketitle

\begin{abstract}
Traditionally, there are three species of classification: unsupervised, supervised, and semi-supervised. Supervised and semi-supervised classification differ by whether or not weight is given to unlabelled observations in the classification procedure. In unsupervised classification, or clustering, all observations are unlabeled and hence full weight is given to unlabelled observations. When some observations are unlabelled, it can be very difficult to \textit{a~priori} choose the optimal level of supervision, and the consequences of a sub-optimal choice can be non-trivial.  A flexible fractionally-supervised approach to classification is introduced, where any level of supervision --- ranging from unsupervised to supervised --- can be attained. Our approach uses a weighted likelihood, wherein weights control the relative role that labelled and unlabelled data have in building a classifier.  A comparison between our approach and the traditional species is presented using simulated and real data.  Gaussian mixture models are used as a vehicle to illustrate our fractionally-supervised classification approach; however, it is broadly applicable and variations on the postulated model can  be easily made.  
\end{abstract}

\section{Introduction}
Broadly, clustering and classification are concerned with assigning labels to observations so that they are partitioned into meaningful groups, or classes. In a model-based setting, classifiers, i.e., functions that map a given observation $\bs x$ to a class label $y$, are constructed based on probability models.  When both $y$ and $\bs x$ are known, we say that the observation is {labeled}. When $\bs x$ is observed and $y$ is missing, the observation is said to be  {unlabeled}.   
We define the data matrix of labelled observations  by 
$\mx_1 = (\bs x_{11} ^{\top},\bs x_{12} ^{\top}, \dots, \bs x_{1n_1} ^{\top}) ^{\top}$
and store the observed class labels in  indicator matrix 
$\mz_1 = (\bs z_{11} ^{\top}, \bs z_{12} ^{\top}, \dots, \bs z_{1n_1}^{\top}) ^{\top}$.  
Herein,  $D_\text{L}$ refers to labelled data comprised of the set  $\{\mx_1, \mz_1\}$.
The unlabelled data,  $D_\text{u}$, are simply the  matrix of unlabelled observations, which we denote by 
$\mx_2 = (\bs x_{21} ^{\top}, \dots, \bs x_{2n_2} ^{\top}) ^{\top}$. Note that $D_\text{u}$ does not include the unknown class labels, which we denote by $\mz_2 = (\bs z_{21}^{\top}, \dots, \bs z_{2n_2}^{\top})^{\top}$.

The task of classification can be performed using a variety of techniques. We briefly mention three `species' of classification here; details of their implementation are provided in Section {\ref{theory}}.
  The first baseline approach is {supervised classification}, wherein labelled data $D_\text{L} = \{(\bs x_{1i}, \bs z_{1i}) \mid i=1,\dots, n_1)\}$ are used to build a classification rule 
   from which to group the remaining unlabelled observations $D_\text{u} = \{\bs x_{2j} \mid j = 1,\dots, n_2\}$. 
 On the contrary, the classifier used in {unsupervised classification} (or {clustering}) relies solely on unlabelled observations. In this framework, 
$\mx_2$ is augmented with latent (unobserved) variables $\mz_2$ and 
objects are grouped, e.g., according to maximum \textit{a~posteriori}
 probabilities. 
Note that clustering is typically used when no labelled points are available (i.e., $D_\text{L}$ is empty). However, to unite these approaches with the generalized method presented herein, clustering will correspond to the case where labelled data (both $\mx_1$ and $\my_1$) are  ignored.

Similar to supervised classification, a {semi-supervised classification} approach includes missing labels $\mz_2$.
 In contrast to supervised classification, semi-supervised classification  makes use of both labelled and unlabelled data, which we denote by $D_\text{o} := D_\text{u} \cup D_\text{L}$.
We highlight that an essential difference between the three species of classification, i.e., supervised, semi-supervised, and unsupervised, is that the classifier is constructed using difference sources of data, i.e., $D_\text{L}, D_\text{o}$, and $D_\text{u}$, respectively. 

Semi-supervised approaches have been employed with success in many applications, including but not limited to \cite{ratsaby1995}, \cite{baluja1998},  \cite{mccallum1998}, \cite{nigam2000}, \cite{mcnicholas2010},  \cite{andrews2011}, and \cite{vrbik14}.  As depicted in Figure \ref{FSCscale}, semi-supervised classification can be viewed as a midpoint between the unsupervised and supervised paradigms. Namely, if we define $\rw$ to be a weight reflecting the relative impact of $D_\text{L}$ and $D_\text{u}$ on building a classifier, a value of $\rw=0.5$ (i.e., where the two sources of data are equally important)  corresponds to semi-supervised classification.  Similarly, the values 0  and 1 coincide with model-based clustering and classification, respectively.  
We propose a method, fractionally-supervised classification (FSC), that enables unlabelled and labelled data to be used to  varying degrees. This approach allows for any intermediate value of $\rw \in [0,1]$ and uncovers a complete spectrum of potential models, of which unsupervised, semi-supervised, and supervised, are  special cases (cf. Figure~\ref{FSCscale}).  
\begin{figure}[!h]
\centering
\begin{tikzpicture}[scale=2.3]
\draw [decorate,decoration={brace,amplitude=5pt,raise=-4pt},yshift=0pt]
(-2,0) -- (2,0) node [black,midway,yshift=1em] {Fractionally-Supervised Classification (FSC)};
\draw [thick] (-2,-.25) -- (2,-.25);
\draw [thick] (-2,-.3) -- (-2,-.2);
\draw [thick] (0,-.3) -- (0,-.2);
\draw [thick] (2,-.3) -- (2,-.2);
\node at (-2,-0.5) {$\rw = 0$};
\node at (0,-0.5) {$\rw = 0.5$};
\node at (2,-0.5) {$\rw=1$};
\node at (-2,-.8) { Unsupervised};
\node at (0,-.8) { Semi-supervised};
\node at (2,-.8) {Supervised};
\end{tikzpicture}
\caption{Value of $\rw$, where $\rw$ represents the relative importance of labelled observations ($D_\text{L}$) versus unlabelled observations ($D_\text{u}$).  Tics represent the special cases of FSC that correspond to existing species of classification. }\label{FSCscale}
\end{figure}

The importance of balancing the respective impact of $D_\text{L}$ and $D_\text{u}$ on a classifier becomes apparent once we consider the following arguments.
 First, although the inclusion of unlabelled data has proven to be beneficial in many classification applications, it is possible that including unlabelled observations may lead to a larger classification error, {e.g.,} when the postulated model is incorrect  \cite[cf.\ ][]{cozman2003,vandewalle2008unlabeled}. 
Further to this, \cite{castelli1996} show that labelled samples are exponentially more valuable than unlabelled samples in reducing classification error when a two-component Gaussian mixture with unknown mixing proportions is considered.  In such cases, it may seem reasonable to assign more weight to labelled observations in the estimation procedure. On the contrary, some situations may benefit by enhancing the role of unlabelled observations; see Section~\ref{sec:iris} for an example of this phenomenon. 

Whatever the case may be, FSC provides a framework that can adjust for the complicated interplay between various sources of data and their relative importance by allowing for any level of supervision between unsupervised and supervised. 
{Derived from the maximum-entropy principle}, our FSC paradigm is based on the weighted likelihood wherein the weights control the contribution of labelled (and unlabelled) observations.  
Herein,  Gaussian mixture models and fixed weights are adopted to illustrate our FSC approach; however,  it is very flexible and can easily be extended to non-Gaussian mixtures and/or different weights. 
 
The remainder of this paper is organized as follows. In Section~\ref{theory}, mixture model-based approaches to discriminant analysis, classification, and clustering are reviewed using notation that will facilitate work described later. The general theory used in the construction of the FSC algorithm is described (Section~\ref{section:WL}) and the model is laid out along with mathematical details (Section~\ref{sec:FSC}). The FSC approach is applied to simulated (Section~\ref{sensitivity}) and real (Section~\ref{sec:App}) data and compared with the three species of classification. The paper closes with some concluding remarks (Section~\ref{sec:con}).

\section{Mixture Models and Classification}
\label{theory}









\subsection{Finite Mixture Models}
\label{sec:fmm}
Finite mixture models have been established as an effective means of classification since they were first used for clustering by \cite{wolfe1965computer}. An example of important early work on classification using mixture models is the work of \cite{scott71}, who outline parameter estimation for Gaussian model-based classification models. They consider equal and unequal component covariance matrices, pointing to previous work by \cite{edwards65} in the former case.

A finite mixture model assumes that a population is comprised of a finite number of subgroups, or components, each following some parametric distribution.  The density of a finite mixture model is a convex linear combination of component densities, given by
\begin{equation}
\mathcal M({\bs x}\mid {\bTheta}) = \sumg \pi_g \phi({\bs x}\mid {\bs\theta}_g),
\label{mixture}
\end{equation}
where $\pi_g>0$ are the mixing proportions such that $\sum_{g=1}^G\pi_g=1$, $\phi(\cdot \mid \bs\theta_g)$ is the $g$th component density parameterized by $\bs\theta_g$, and $\bTheta=(\pi_1, \dots, \pi_G, \bs\theta_1,\dots, \bs\theta_G)$ is the vector of mixture parameters.

\subsection{Model-Based Classification}\label{sec:class}
Consider a sample of $N$ $d$-dimensional observations, independently drawn from (\ref{mixture}). 
Suppose that we know the labels, i.e., the components of origin, for a subset of the observations so that an ${n\times d}$ matrix 
 $\mx = (\mx_1^{\top}, \mx_2^{\top})^{\top}$ comprises the labelled observations $\mx_1 = (\bs x_{11}^{\top}, \bs x_{12}^{\top}, \dots,\bs  x_{1n_1}^{\top})^{\top}$ and the unlabelled observations $\mx_2 = (\bs x_{21}^{\top}, \bs x_{22}^{\top}, \dots, \bs x_{2n_2}^{\top})^{\top}$, where $n_1 +  n_2=N$.  Suppose that $\mx_i$ has associated component indicator matrix $\mz_i$ = $(\bs z_{i1}^{\top}$,$ \bs z_{i2}^{\top}$, \dots, $\bs z_{in_i}^{\top})^{\top}$, where $\bs z_{ij} = (z_{j1}^{(i)}, z_{j2}^{(i)}, \dots, z_{jG}^{(i)})$ with elements
\begin{equation}
\label{zig}
z_{jg}^{(i)} =
\begin{cases}
1, & \text{if }\bs x_{ij}\text{ arises from component }g,\\
0, & \text{otherwise,}
\end{cases}
\end{equation}
for 
{$i=1,2$} and 
$j=1,\ldots,n_i$.
As defined in the previous section, we use the notation $D_\text{L} = \{\mx_1, \mz_1\}$, $D_\text{u} = \{\mx_2\}$, and $D_\text{o} = D_\text{L} \cup D_\text{u}$.

Semi-supervised classification  maximizes the {observed-data likelihood function} of $\bTheta$ given the observed data $D_\text{o}$:
\begin{align}
\label{osbclass}
\Lsemi
&= \prod_{j=1}^{n_1}\prodg[{\pi_{g}}  \phi(\bs x_{1j}\mid\bs\theta_g)]^{\zigone} 
\times\prod_{j'=1}^{n_2}\sum_{h = 1}^H \pi_{h} \phi(\bs x_{2j'}\mid\bs\theta_h),
\end{align}
where $G$ is the number of classes amongst the labelled observations and $H\geq G$ is the number of components in the fitted model. 
To maximize (\ref{osbclass}) with respect to $\bTheta$, we make use of the expectation maximization (EM) algorithm \citep{dempster77}. The EM algorithm 
 uses 
 the so-called 
complete-data, which we denote by $D_\text{c} := \{\mx, \mz\}$, where $\mz = (\mz_1^\top, \mz_2^\top)^\top$. 
The {complete-data likelihood function} for $\bTheta$ given the complete-data $D_\text{c}$ is given by
\begin{align}
\label{compclass}
\lik{c}{D_\text{c}}
&=\prodjone \prodg [\mixx{1j}]^{\zigone}  \times\prod_{k=1}^{n_2}\prod_{h = 1}^H [\pi_h \phi(\bs x_{2k}\mid\bs\theta_h)]^{z_{kh}^{(2)}},
\end{align}
where $H\ge G$.  As is common practice in classification problems, we  restrict ourselves to $H=G$ hereafter.  This is equivalent to assuming that all classes represented amongst the unlabelled observations are also represented amongst the labelled observations. Furthermore, one should note that relaxing this assumption, i.e., allowing $H>G$, requires only very minor modifications to the methodology described hereafter. Further discussion about the number of components, in this context, is given in Section~\ref{sec:GH}.

The EM algorithm iterates through an expectation step (E-step) and a maximization step (M-step) until a stopping criterion is satisfied. Starting with an initial estimate for $\bTheta$, denoted by $\bTheta^{(0)}$, the EM algorithm proceeds as follows, where $(t)$ indicates the iteration~$t$.
\begin{subequations}\label{EM2}
\begin{align}
  \label{estep1}
& \text{\textbf{E-step}}  \text{ Update } \zighattwo = \frac{\mixt{2j}{t}}{\sumg \mixt{2j}{t}} \text{ for }  j = 1, \dots, n_2.
 \\
   \label{mstep1}
& \text{\textbf{M-step}} \text{ Update } \bTheta\ptp= \underset{\bTheta}{\operatorname{argmax}} \  \sumi \sumj \sumg {\zighat}\log   \big[\mixt{ij}{t}\big].\\
\label{stopstep}
& \text {If converged, stop. Otherwise, set } t \leftarrow t+1 \text{ and repeat.}
\end{align}
\end{subequations}
For ease of notation, set $\hat z_{jg}^{(1)} = z_{jg}^{(1)}$ and 
\begin{equation}
Q(\bTheta \mid \bTheta\pt) = E_{p(\mz_2 |\mx)}\left[\ell_{c}(\bTheta \mid {D_\text{c}}) \mid D_\text{o},  \bTheta\pt \right],
\label{q}
\end{equation}
where $\ell_{c}(\bTheta \mid D_\text{c}) \coloneqq \log \lik{c}{D_\text{c}}$.
Now, (\ref{estep1}) is equivalent to calculating (\ref{q}) based on the current values of the parameter estimates $\bTheta\pt$(E-step) and (\ref{mstep1}) is equivalent to maximizing (\ref{q}) with respect to ${\bTheta}$. 

\subsection{Model-Based Discriminant Analysis}\label{sec:DA}
Model-based discriminant analysis is a supervised approach to classification that uses only labelled data to build a classifier.   We consider the quadratic discriminant analysis (DA) rule that determines the class of an unlabelled subject $j$ according to 
\begin{equation}\label{DArule}
\zighattwo = \frac{\mixhat{2j}}{\sumg \mixhat{2j}}.
\end{equation}
Specifically, $\bs x_{2j}$ is classified into component $g$ if 
$\zighattwo > \hat{z}_{jk}^{(2)}$ 
for all $k \neq g$,
and $\hat\bTheta = \{\hat{\pi}_g, \hat{\bs\theta}_g\mid {g=1,\dots,G}\}$ are the estimates found by maximizing 
\begin{equation}
\label{DA}
\Lda = \prodjone \prodg[\mixx{1j}]^{z_{jg}^{(1)}}.
\end{equation}
Unlike model-based classification, $\hat\bTheta$ can be found using traditional likelihood maximization techniques. 
More specifically,  maximizing (\ref{DA}) with respect to $\bTheta$ leads to the plug-in estimates,
\begin{equation*}
\begin{split}
\hat{\pi}_g  =& \frac{\sum_{j=1}^{n_1} \zigone}{n_1},\qquad
\hat{\bm\mu}_{g} =  \frac{\sumjone \bs x_{1j} \zigone }{\sumjone \zigone},\qquad
 \hat{\covar}_{g} = \dfrac{\sumjone \zigone(\bs x_{1j}-{\bm\mu}_{g})(\bs x_{j1}-{\bm\mu}_{g})^{\top}}{\sumjone \zigone}, 
\end{split}
\end{equation*}
for $g=1,\dots, G$.

\subsection{Model-Based Clustering}\label{sec:clustering}
\label{MBCL}
Model-based clustering is an unsupervised approach and so uses no prior knowledge of class labels.  Consequently, this method aims at maximizing the observed likelihood given by
\begin{align}
\label{oclustlike}
\Lclust  &= \prodjtwo \sumg \mixx{2j}.
\end{align}
This is accomplished by augmenting the data $D_\text{u}$ with missing labels $\mz_2$ and using the EM 
 algorithm 
 in \eqref{estep1}--\eqref{stopstep}. As mentioned previously, $\mx_1$ and $\mz_1$ are treated as empty.

\section{Weighted Likelihood}
\label{section:WL}

In his seminal paper, \cite{stein1956} explored the notion of borrowing information from samples drawn from populations other than the population of direct interest. The author changed conventional estimation approaches when he showed that the maximum likelihood estimators (MLEs) for the means of independent Gaussian populations with known variance are inadmissible when the number of populations exceeds two. This phenomenon led to a number of alternative estimators that dominate the MLE, including but not limited to the celebrated James-Stein estimator \citep{james1961}, the modified James-Stein estimator \citep{berry1994}, and the minimax estimators presented by \cite{baranchik1970} and \cite{strawderman1973}. These Stein-type estimators are closely connected to the weighted likelihood. In fact, when the weights are appropriately defined, the James-Stein estimator can be written as a special case of the weighted likelihood estimator \citep{wang2006}.
 
Another closely connected forerunner to the weighted likelihood is the relevance weighted likelihood \citep[REWL;][]{hu1994, hu1997,hu2001}. The REWL  borrows information from distributions thought to be similar to the distribution of inferential interest to produce better estimates, in terms of mean squared error, of the target distribution. The asymptotic properties of this class of estimators generalize the work of \cite{wald} for the traditional MLE and are presented for the parametric case by \cite{hu1997}.

Not long after, the weighted likelihood was formulated. Subsequently, it has been theoretically solidified and expounded upon by several authors  \citep[e.g.,][]{wang2001, huZidek2002, wang2004, wang2005, wang2006, plante2008, plante2009}.  The weighted likelihood adopts a different paradigm than the REWL, which assumes information about $\bTheta$ increases as the number of populations grows \citep[cf.][]{hu1997}. In contrast, the asymptotic theory of weighted likelihood estimators assumes that the number of populations is fixed as sample size tends to infinity \citep[cf.][]{wang2004}. 
The derivation of both the non-parametric and parametric versions of their weighted likelihood are detailed by \cite{huZidek2002}.
 
In the weighted likelihood framework, it is assumed that data come from $m$ populations having similar, but not necessarily identical, distributions. To be more specific, let $\my_i = (\bs y_{i1}^{\top}, \dots,\bs y_{in_i}^{\top})^{\top}$, $i=1,\dots,m$, where $\bs y_{ij}$ are independent random vectors with probability density functions (PDFs) $f_i(\cdot \mid \bTheta_i)$. Let $\mat{Y} = \{\my_i \mid i=1, \dots, m\}$
and assume $f_2(\cdot\mid\bTheta_2), \dots , f_m(\cdot\mid\bTheta_m)$ are similar to  our primary PDF of interest $f_1(\cdot\mid\bTheta_1)$, then the weighted likelihood takes the form
\begin{equation}
\mathcal L_w(\bTheta_1\mid \my) = \prodim \prodj f_1(\bs y_{ij} \mid \bTheta_1)^\ai,
\label{WL}
\end{equation}
where $\rw_1, \rw_2, \dots, \rw_m$ are likelihood weights.
 Herein, we assume positive-valued weights and fix $\rw_1 + \dots + \rw_m = 1$; however, this restriction could be relaxed \citep[cf.][]{huZidek2002}. 
Naturally, the maximum weighted likelihood estimator (MWLE) is found by maximizing the weighted likelihood presented in (\ref{WL}). Asymptotic properties for the class of MWLE estimators are given by \cite{wang2004}.  

%
%
%
%
%

\section{Fractionally-Supervised Classification}
\label{sec:FSC}

\subsection{The Model}
\label{sec:mod}
Our approach can be described by adopting  the weighted likelihood defined in (\ref{WL}) with two populations (i.e., $m=2)$. 
Suppose $\my_1$ are the data pairs provided by labelled data $D_\text{L}=\{(\bs x_{1j}, \bs z_{1j}) \mid j=1,\dots, n_1)\}$
 drawn from $f_1( \cdot \mid \bTheta_1)$, the distribution function of inferential interest.  Let $\my_2 = \{(\bs x_{2j'}, \bs z_{2j'}) \mid j'=1,\dots, n_2)\}$, where  $\mz_2 = (\bs z_{21}, \dots, \bs z_{2n_2})$ are the missing labels associated with  unlabelled observations $(\bs x_{21}, \dots, \bs x_{2n_2} )$.  We view the samples $\my_1 $ and $\my_2$ as being drawn from similar populations. 
The weighted likelihood is therefore given by
\begin{equation}
\wLc=  \prodi \fonexzi^{\ai }=  \prodi \prodj \fcompj{i} ^{\ai },
\label{cwl}
\end{equation}
where 
$$\fcompj{i} = \prodg[\mixx{1j}]^{z_{jg}^{(i)}}$$ and $\rw_1 + \rw_2 = 1$.  The associated MWLE is found by maximizing \eqref{cwl}.

The MWLE described above can also be arrived at by adding weights to the traditional likelihood used in semi-supervised classification.  More specifically, we  define a weighted version of \eqref{osbclass}, given by
\begin{align}
\label{eq:wl}
\begin{split}
\wL& = \left[\Lda\right]^{\aone} \times \left[\Lclust\right]^{\atwo}\\
=& \left[\prod_{j=1}^{n_1}\prodg[{\pi_{g}}  \phi(\bs x_{1j}\mid\bs\theta_g)]^{\zigone} \right]^{\aone} 
\times\left[\prod_{j'=1}^{n_2}\sum_{g = 1}^G \pi_{g} \phi(\bs x_{2j'}\mid\bs\theta_g)\right]^{\atwo}.
\end{split}
\end{align}
By maximizing \eqref{eq:wl} using the EM algorithm, 
it follows that the complete-data log-likelihood used in the M-step is equivalent to \eqref{cwl}.  In our opinion, this is the most natural construction for our weighted likelihood; however, it does not precisely fit within either the weighted likelihood or REWL paradigms.  Noting that this discrepancy does not change the general conclusion, we henceforth use `weighted likelihood' to refer to \eqref{eq:wl} and define our FSC estimator (FSCE) as
\begin{equation}
\underset{\bTheta}{\operatorname{argmax}} \ \wL,
\label{owl}
\end{equation}
where $(\rw_1, \rw_2)=(\ac, 1-\ac)$. 
{$\text{FSC}_{\awhat}$} is used to denote the FSCE with $\rw_c={\awhat}$. 
{ When ${\awhat}$ corresponds to supervised, unsupervised, or semi-supervised classification, the subscript is replaced with `DA', `Clust', or `Class', respectively.  }
The FSCE is found using the EM algorithm, as summarized below:

\noindent \textbf{E-step}: 
{Given $\bTheta = \bTheta\pt$} at iteration $t$, compute
$$ \hat z_{jg}^{(2)} = \frac{\pi\pt_{g} \phi_{g}({{\bm x}_{2j}}\mid {\bs\theta}\pt_{g})}
{\sumg \pi\pt_{g} \phi_{g}({{\bm x}_{2j}}\mid {\bs\theta}\pt_{g})},$$ 
for $j = 1, \dots n_2, g = 1, \dots, G$.   
Recall that $\hat z_{jg}^{(1)}$ are  defined to be the known labels  $z_{jg}^{(1)}$.\\[5mm]
\textbf{M-step}:  

Update $\pi_{g}\pt$ by maximizing (\ref{cwl}) with respect to $\pi_g$, which leads to
$$ {\pi_{g}}\ptp =\dfrac{\sumi \sumj{\ai  {\hz}}}{\sumi \sumj \sumg {\ai  {\hz}}}.$$ 

Update $\bm\mu_g\pt$  by maximizing (\ref{cwl}) with respect to $\bm\mu_g$, which leads to
$$\bm\mu_{g}\ptp = \frac{\sumi \sumj \bs x_{ij}\ai \hz}{\sumi \sumj \ai \hz }.$$

Update $\bs\Sigma_g\pt$  by maximizing (\ref{cwl}) with respect to $\bs\Sigma_g$, which leads to
$$\bs\Sigma_{g}\ptp = \dfrac{\sumi\sumj \ai \hat z_{jg}^{(i)}(\bs x_{ij}-{\bm\mu}_{g}\ptp)(\bs x_{ij}-{\bm\mu}_{g}\ptp)^{\top}}{\sumi \sumj \ai \hat z_{jg}^{(i)}}.$$

We initialize our FSC algorithm using $k$-means clustering \citep{hartigan1979} and use a lack of progress stopping criterion.  To be specific, the EM algorithm is terminated when the difference between successive weighted log-likelihoods, $\wl \coloneqq \log\{\wL\}$, is less than some small $\epsilon$.  The analyses herein use
$\wl^{(t)} - \wl^{(t-1)}  < 1 \times 10^{-5}.$

\subsection{Related Work}\label{rw}
In work on a semi-supervised classification framework, \cite{sokolovska2008} present a weighted likelihood 
 estimator  
 that was shown to perform at least as well as the maximum likelihood estimator (i.e., the estimator found in the supervised setting) under certain conditions.  The information provided by the unlabelled data is incorporated through a weight given by the ratio of the densities, 
\begin{equation*}
\frac{g(\bs x \mid \bs\theta')}{g(\bs x \mid  \bs\theta)},
\end{equation*}
or functions thereof, where $\bs\theta'$ and $\bs\theta$ are the MLEs found using the unlabelled and labelled data, respectively. This idea was later extended to the density ratio estimation-based semi-supervised  (DRESS) estimators of \cite{kawakita2014}, which use modified weights.  

It is important to note that the objectives of FSC are fundamentally different from those of the aforementioned density ratio estimation methods. FSC is a paradigm that allows classification to be carried out with any level of supervision, running from unsupervised to supervised, with semi-supervised classification representing the mid-point; density ratio estimation methods, however, deal with improvements to supervised classification.
Put another way,  the DRESS estimation procedure uses unlabelled data  solely to calculate weights to be assigned to labelled observations $\mx_1$; hence, $\mx_2$ is not explicitly used in building a classifier.  In FSC, the entire set of observed data, i.e., $D_\text{o} =  D_\text{L} \cup D_\text{u}$, is used to estimate $\bs \Theta$ and weights are required for both $\mx_1$ and $\mx_2$.  Furthermore, the DRESS estimator defines different weights for each individual point $\bs x_{1j}, j = 1,\dots,n_1$.  It could be that observation-specific weights, say $\omega_{ij}, i = 1,2$ and $j=1,\dots, n_i$, might advantageously be employed in the FSC framework; however,    such investigations are beyond the scope of this paper.  

\cite{nigam2000} present a weighted extension to the EM algorithm, termed EM-$\lambda$, which was explored in  the context of text classification problems.  Similar to FSC, EM-$\lambda$ uses a weighting factor $\lambda \in [0,1]$ to moderate the contribution of unlabelled data on building a classifier. 
Unlike FSC, EM-$\lambda$ fails to  provide a corresponding weight for labelled observations.  As a consequence, labelled data are always given full weight in parameter estimation and the effect of unlabelled data can only be weakened.   On the other hand, FSC moderates the effect of unlabelled data in both directions; hence, weights can be used either to enhance or decrease the effect of  $D_\text{L}$  (and  $D_\text{u}$) when building a classifier.  An example of where this distinction becomes critical is given in Section \ref{sec:iris}, where we show that enhancing the role of unlabelled observations can lead to 
an improvement in the resulting classification.

\subsection{ Weight Specification}
\label{weights}

{In the context of FSC, weights are used to emphasize (or soften) the  role of labelled observations ($D_\text{L}$) and unlabelled observations ($D_u$) in parameter estimation.  To tie this into the WL framework, one might say $\omega_i$ reflects how well  sample $\my_i$ represents the population of inferential interest.}
Selecting the  weight $\ac$ is a nontrivial problem. Because  the FSCE is found by maximizing the WL, one may be tempted to do the same for $\ac$.  
If this were our goal, $\ac$ would always be selected to be a boundary point because the weighted log-likelihood is a linear function of $\ac$.  
Specifically, the weighted log-likelihood would be maximized with $\ac$ = 0 if 
 $\Ldahat < \Lclusthat$ or $\ac = $ 1 if $\Lclusthat < \Ldahat$.


Some progress has been made towards specifying adaptive, i.e., data-dependent, weights.
 For example,
\cite{wang2005} offer an adaptive approach that relies on a {leave-one-out} cross-validation technique.  They choose their weights  to
minimize a measure of discrepancy given by
\begin{equation}\label{dis}
D(\abold) = \sum_{j=1}^{n_1} (\x_{i1} - \tilde{\bm\mu}\kout)^2,
\end{equation}
  where $\tilde{\bm\mu}\kout$ is the  WLE of the mean found when the $k$th observation has been dropped from the sample.    For the examples considered therein, $\tilde{\bm\mu}\kout$ can be expressed as a linear combination of the MLEs for each population and the optimal weights can be found analytically. This is not the case for finite mixture models.  We could conceivably calculate \eqref{dis} for a set of candidate weights and select the value that obtained the lowest discrepancy score; however, this technique would come at a high computational cost.

\cite{huZidek2002} offer another data-driven alternative based on a weighted log-likelihood ratio.  It should be noted that they aim at simultaneously estimating $f_1( \cdot \mid \bTheta_1),\dots, f_m( \cdot \mid \bTheta_m)$. Consequently, $\bTheta_1, \dots, \bTheta_m$ appear in their version of the WL, which differs from \eqref{WL}.  Keeping this discrepancy  in mind, we first introduce the procedure as described therein, then adapt it to suit the FSC paradigm. 
The authors define $\hat\bTheta_{\abold} = (\hat\bTheta_{1\am{1}}, \dots \hat\bTheta_{m\am{m}})$ as the WLE of $\bTheta = (\bTheta_1,\dots, \bTheta_m)$ based on weights $\abold = (\rw_1, \dots, \rw_m)$, and define the optimal weight vector to be
\begin{equation}\label{aopt}
\aopt(\bTheta) = \underset{\abold}{\operatorname{arg \ max}} \sum_{i=1}^{m} E\left[\int f_i(\x \mid \bTheta_i) \log \{f_i(\x \mid \hat\bTheta_{i\rw_i})  \} d\x \right].
\end{equation}
In practice, the authors suggest replacing $\bTheta_i$  by a reasonable estimate (such as the MLE of $\bTheta_i$) and calculating \eqref{aopt} using a second-order Taylor series approximation.

Drawing from this idea, we could select $\ac$ to maximize
\begin{equation}\label{aoptours}
E \left[ \int \ftheta{1} \log \fhatc{1} d\x +  \int \ftheta{2} \log \fhatc{2} d\x \right].
\end{equation}
It is easy to show that \eqref{aoptours}  can be rewritten as
\begin{align*}
E\left[C - \text{KL}\left(\fonehat \mid \mid \fhatc{1} \right) - \text{KL}\left(\ftwohat \mid \mid \fhatc{2} \right)\right],
\end{align*}	
where $$C = \int \fonehat \log \fonehat d\x + \int \ftwohat \log \ftwohat d\x$$ and $\text{KL}(g\mid\mid f)$ is the Kullback-Leibler (KL) divergence \citep{kullback1951} between probability distributions $f$ and $g$.  Because $C$ is a constant with respect to $\ac$, we can define an optimal  weight to be
\begin{equation}\label{aoptadj}
\aklopt{} = \underset{\ac}{\operatorname{arg\  min}} \
\sum_{i=1}^2 E \left[\text{KL}\left(\ftheta{i} \mid \mid \fhatc{i} \right) \right].
\end{equation}
In the development of the famous  Akaike information criterion (AIC), an asymptotically unbiased estimator of the relative expected KL divergence was shown to be $\ell(\hat\bTheta \mid D) - K$, where $\ell(\hat\bTheta \mid D)$ represents the log-likelihood of $\hat\bTheta$ given data $D$ and $K$ is the number of free model parameters \citep{akaike1973information}.
Because each model requires the same number of free parameters to be estimated, $\aklopt{}$ can be chosen to maximize 
\begin{equation}\label{AIC}
\log(\Ldahat) + \log(\Lclusthat) =  \log (\Lsemihat),
 \end{equation}
 i.e., the unweighted log-likelihood for semi-supervised classification; see~\eqref{osbclass}.
In other words,  using this strategy for selecting $\ac$ is equivalent to doing model-based classification.   
Because  our aim is to interpolate between the three species of classification, it is clear that alternative selection techniques are needed.
 
For lack of an adequate solution, we investigate a range of $\ac$ values in the upcoming analyses (Sections~\ref{sensitivity} and~\ref{sec:App}). 
Because the primary goal of this procedure is to classify observations into their group of origin, the `best' value of $\ac$ may be considered to be the one that attains a partition of the data closest to the truth. 
To gauge the the efficacy of each model, we use the adjusted rand index \citep[ARI;][]{hubert85}. The ARI  measures the pairwise agreement between two partitions while accounting for chance agreement. 
Note that an ARI of 1 corresponds to perfect classification, and the expected value of the ARI under random classification is 0 \cite[cf.][]{steinley2004properties}.  
Through these applications we show that the optimal choice for $\ac$ (in terms of ARI) differs from case to case and infrequently  corresponds to one of the three species.  

In Section~\ref{sensitivity}, we consider examples for which the complete-data are available, i.e., where the class labels are known for all observations. We proceed by randomly selecting observations to treat as unlabelled, then we check that all classes are represented in the labelled and unlabelled data.  

{\subsection{Number of Components}\label{sec:GH}
As mentioned previously, the number of fitted components is typically set equal to the number of class labels amongst the labelled data, i.e., $H=G$. It may happen, however, that one or more of the underlying classes are not represented amongst the labelled observations, i.e., we would need $H>G$. The risk of this happening may be particularly high when labelled observations are scarce or there are many classes.  
Our recourse in this situation is to fit a set of candidate models, e.g., $H= G, \dots, G+3$, and choose the number of components based on some model selection criterion such as the BIC \citep{schwarz1978estimating} or ICL \citep{biernacki2000assessing}. In fact, this is essentially the approach that is typically taken in the unsupervised paradigm.}

{Note that it is possible that the number of classes in the labelled data is actually more than the true number of classes. While perhaps an atypical situation, it is nonetheless possible and can occur, e.g., when labels correspond to some lower level substructure within a cluster. Within our approach, there is no mechanism to directly fit $H<G$ without ignoring some or all of the known class labels. However, one can fit an $H$-component model and consider \textit{a~posteriori} cluster merging as a secondary analysis \citep[see, e.g.,][]{baudry10,hennig10}. While these considerations are possible under the FSC framework, for the sake of simplicity, we confine the analyses in Section~\ref{sensitivity} to cases where $H=G$. } 

\section{Simulation Studies}\label{sensitivity}

To test how the performance of the FSC approach is affected by the specification of $\ac$, 
we fit FSC for a set of candidate weights
$\awhatvec = 
\{0, 0.1, \dots, 0.9, 1\}$.  For each simulation, $r$, the component of origin for  $(100-p)$\% of observations is hidden from the classification procedure so that $\bs\mx_{(rp)} = \left\{\mx_{1(rp)},\mx_{2(rp)}\right\}$ comprise the labelled data
 $\mx_{1(rp)} = (\bs x_{1(r_1)}^{\top},  \dots,\bs  x_{1(r_{n_1})}^{\top})^{\top}$ 
 and the unlabelled data
 $\mx_{2(rp)} = (\bs x_{1(r_{n_1+1})}^{\top},  \dots,\bs  x_{2(r_{n})}^{\top})^{\top}$. In a similar fashion, let
 $\mz_{1(rp)} = (\bs z_{1(r_1)}^{\top},  \dots,\bs  z_{1(r_{n_1})}^{\top})^{\top}$ 
and 
 $\mz_{2(rp)} = (\bs z_{1(r_{n_1+1})}^{\top},  \dots,\bs  z_{2(r_{n})}^{\top})^{\top}$  
denote the component matrices for the labelled and  unlabelled data, respectively.  
We consider nine `splits' on 100 data sets, i.e., $\{\bs\mx_{(rp)} \mid p = 10, 20, \dots, 90\ , r = 1,2, \dots, 100\}$. 
The observed data are denoted by $D_{rp} := \{\bs\mx_{(rp)}, \mz_{1(rp)}\}$. 

 We compare the clustering results of  $\text{FSC}_{\awhat}$ ($\awhat \in \awhatvec)$ with 
 the `best' fitted FSC solutions in terms of the ARI. More specifically, we let $\text{ARI}_r(\awhat, D_{rp})$ be the ARI associated with $\frhat$ fitted to data $D_{rp}$ and compute its average value using
\begin{equation}\label{avgari}
\avgari =  \sum_{r=1}^{100} \frac{\text{ARI}_r(\awhat, D_{rp})}{100}.
\end{equation}
 The optimal weight obtained by maximizing \eqref{avgari}  is given by
$\ariopt = \underset{\awhat   \in  \awhatvec}{\operatorname{arg\ max}}
\ \avgari.$
The fitted FSC model with $\ac = {\ariopt}$ is denoted by $\fari$.  As previously mentioned, the special cases of FSC, namely, $\text{FSC}_{0}$, $\text{FSC}_{0.5}$, and $\text{FSC}_{1}$, are designated by $\fclust, \fclass$,  and $\fda$, respectively. Note that $\fclust$ ignores, rather than treating as unlabelled, the labelled observations.

\subsection{Simulation 1}\label{sec:sim1}
\begin{figure}[!h]
\vspace{-0.12in}
\begin{center}
\includegraphics[width=0.3754\textwidth]{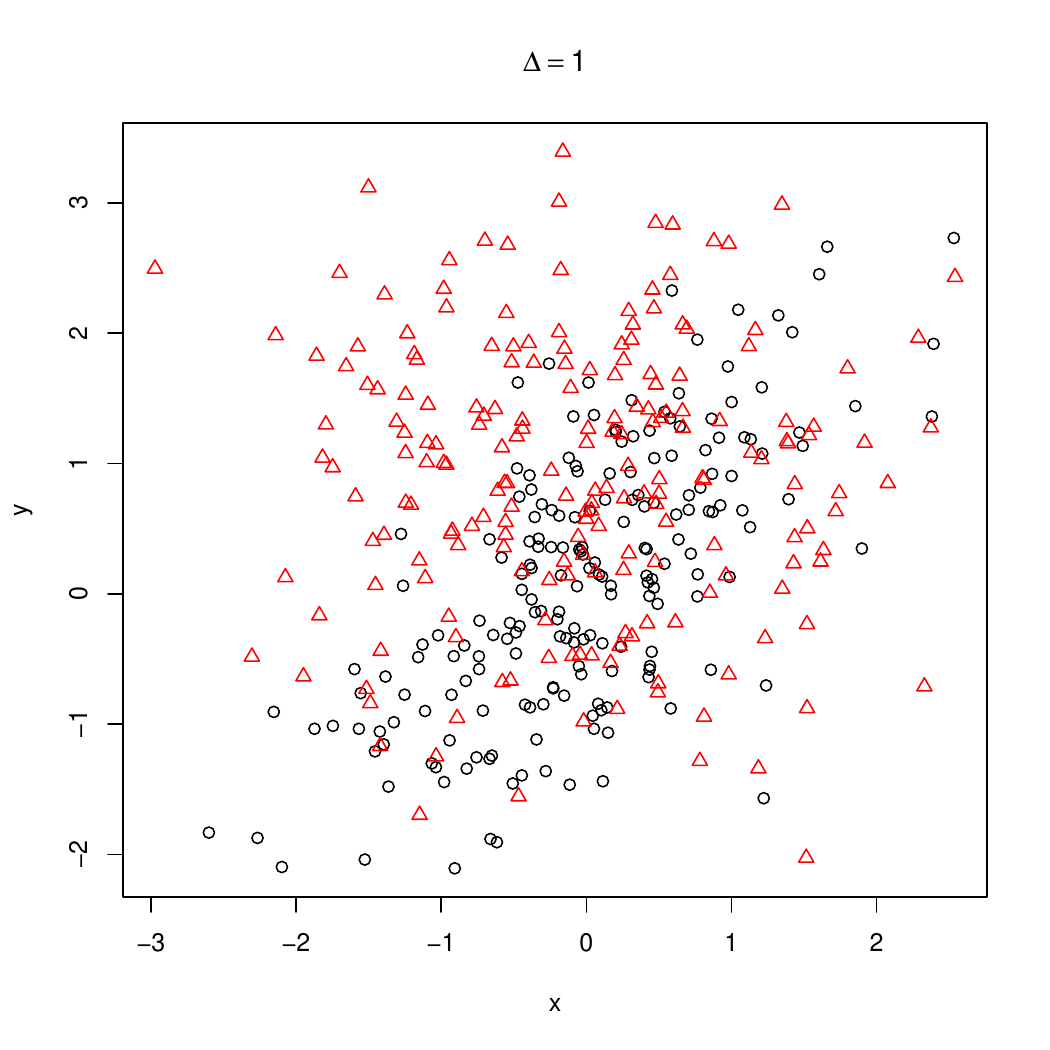} \qquad
\includegraphics[width=0.3754\textwidth]{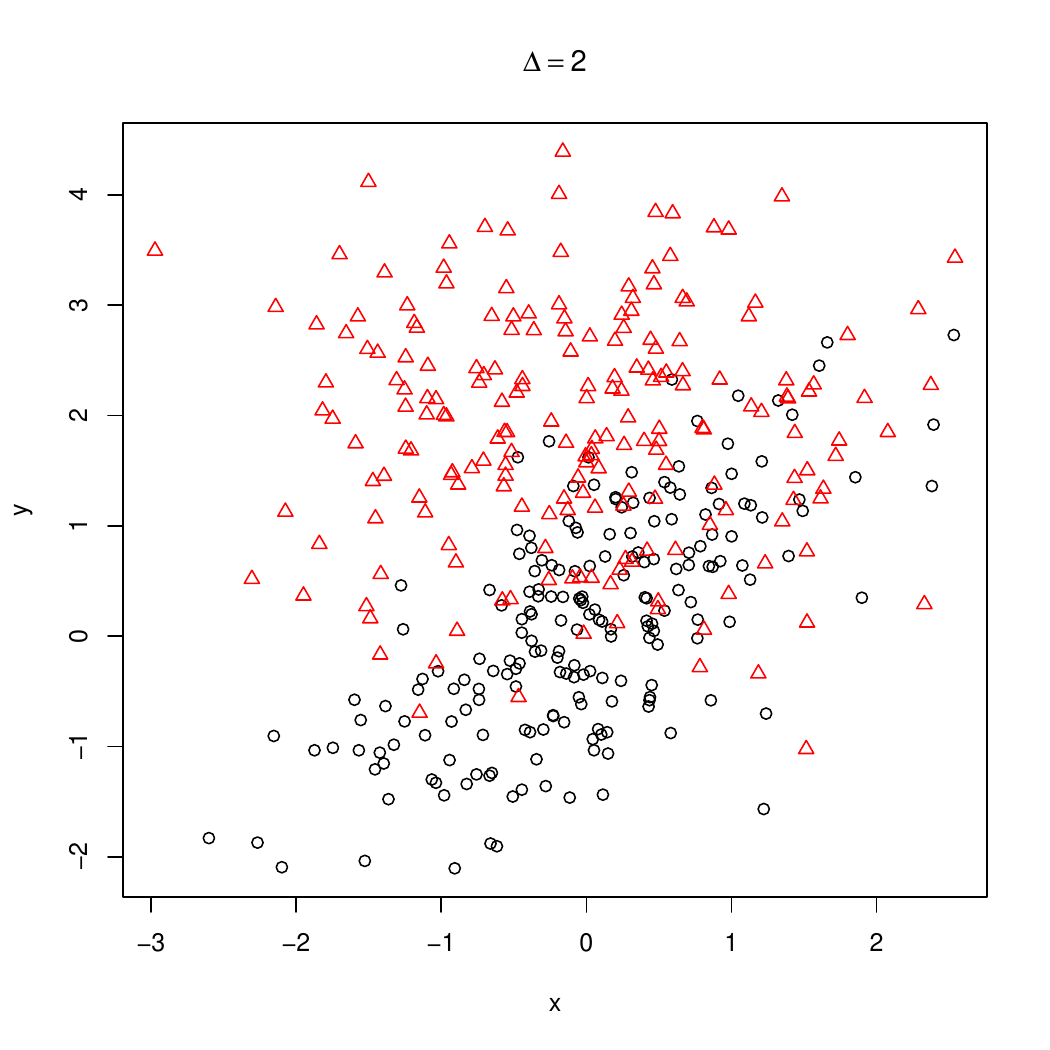}
\includegraphics[width=0.3754\textwidth]{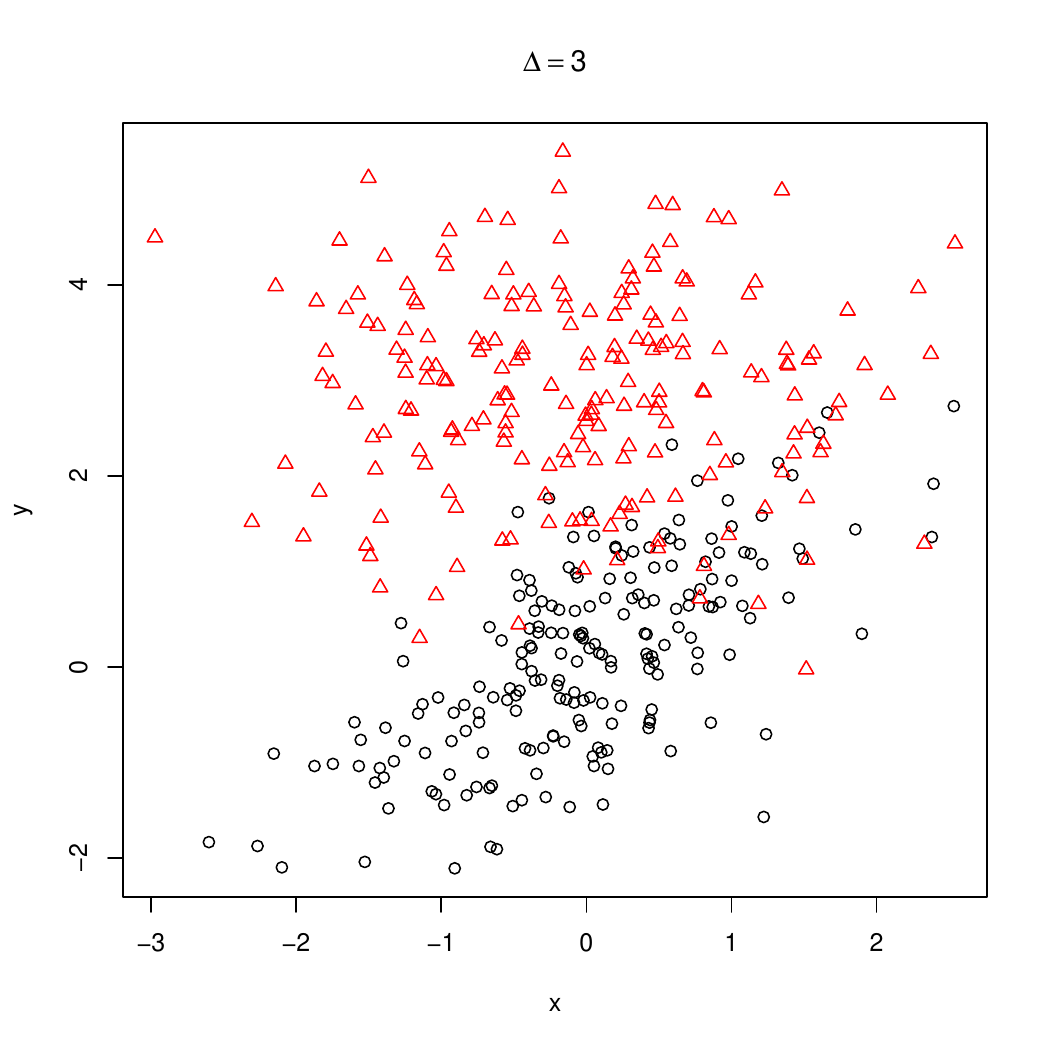} \qquad
\includegraphics[width=0.3754\textwidth]{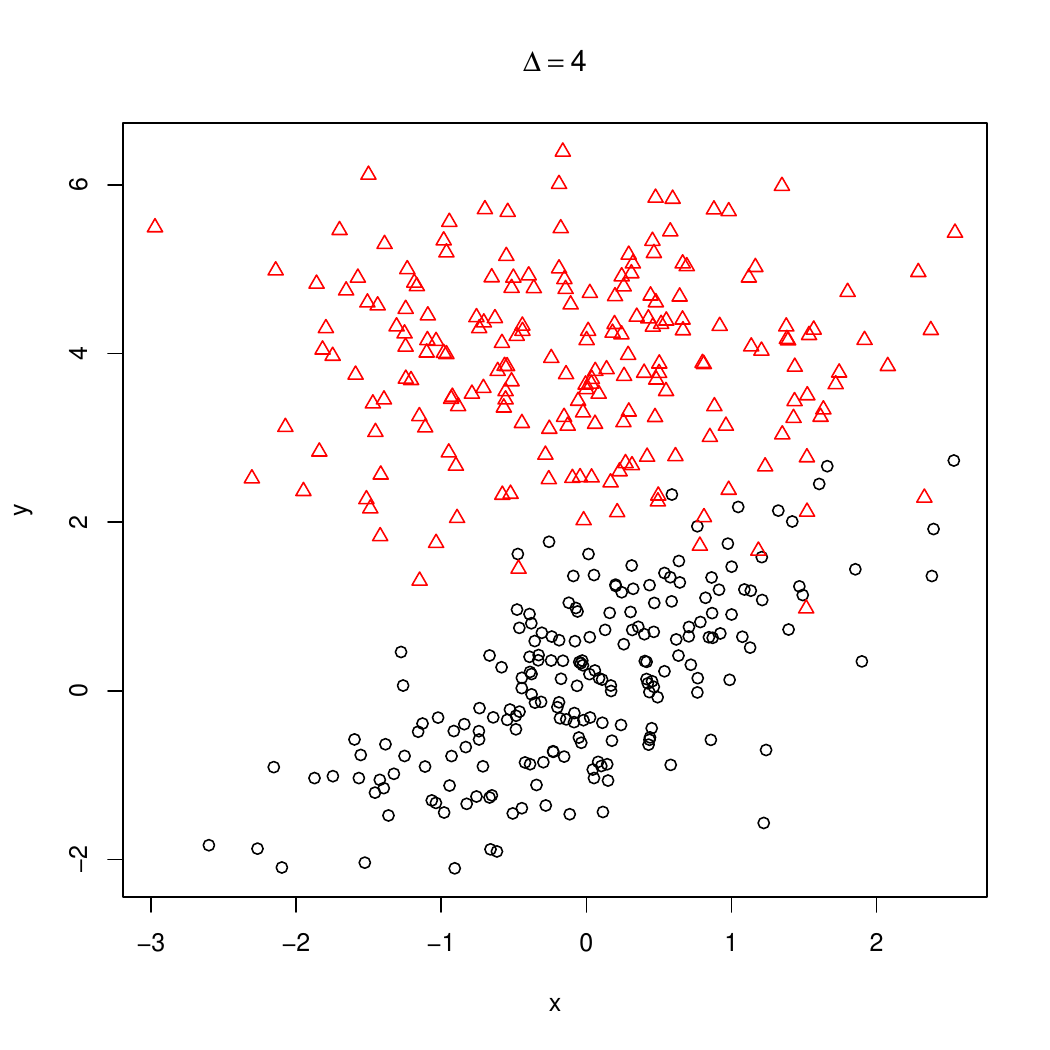}
\includegraphics[width=0.3754\textwidth]{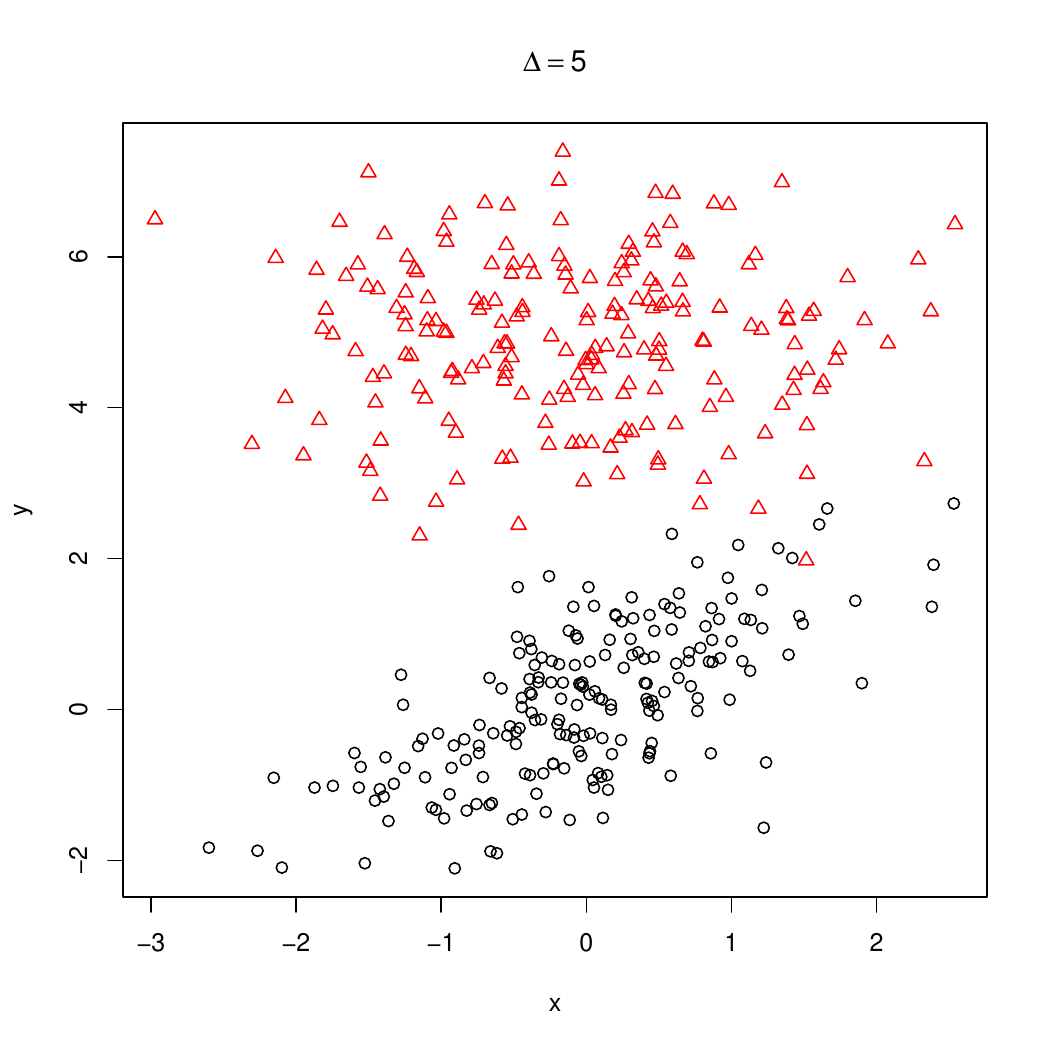}
\end{center}
\vspace{-0.12in}
\caption{ Typical data sets simulated from \eqref{eq:sim}, illustrating varying degree of separation, where $\Delta$ indicates the Euclidean difference along the $y$-axis between  group one ({plotted using black circles}) and group two ({plotted using red triangles}).}
\label{qsim}
\end{figure}

The data under consideration are simulated using a two-dimensional two-component mixture model with varying degrees of overlap.  We denote the Euclidean distance between  group means by $\Delta$, so that 
we have a finite mixture defined as
\begin{equation}\label{eq:sim}
\pi \phi(\bs x \mid  \bs\mu_1, \bs\Sigma_1) + (1-\pi)\phi(\bs x \mid  \bs\mu_2, \bs\Sigma_2),
\end{equation}
where  $\pi = 0.5$, $\bs\mu_1 = (0, 0)^{\top}, \bs\mu_2 = (0,\Delta)^{\top}$, and 
\begin{equation*}
\bs\Sigma_1= \left( \begin{array}{ccc}
1 & 0.7\\
0.7 & 1\end{array} \right),\qquad
\bs\Sigma_2= \left( \begin{array}{ccc}
1 & 0\\
0 & 1\end{array} \right).
\end{equation*}
We consider 100 data sets ($r=1,\dots,100$)  drawn from \eqref{eq:sim} with 300  observations ($n_1=n_2=150)$  for each $\Delta \in \{1,2,\dots,5\}$, and $p \in \{10, 20, \dots,  90\}$. 
In this section, we use an extra data subscript to indicate the group separation, i.e., $\dataD$ is used instead of $D_{rp}$.
The FSCE is found for each simulation using the algorithm summarized in Section \ref{sec:mod}. Data from a typical simulation, for the different values of $\Delta$, are illustrated in Figure~\ref{qsim}.

\begin{figure}[!h]
\vspace{-0.12in}
\begin{center}
\includegraphics[width=0.49\textwidth]{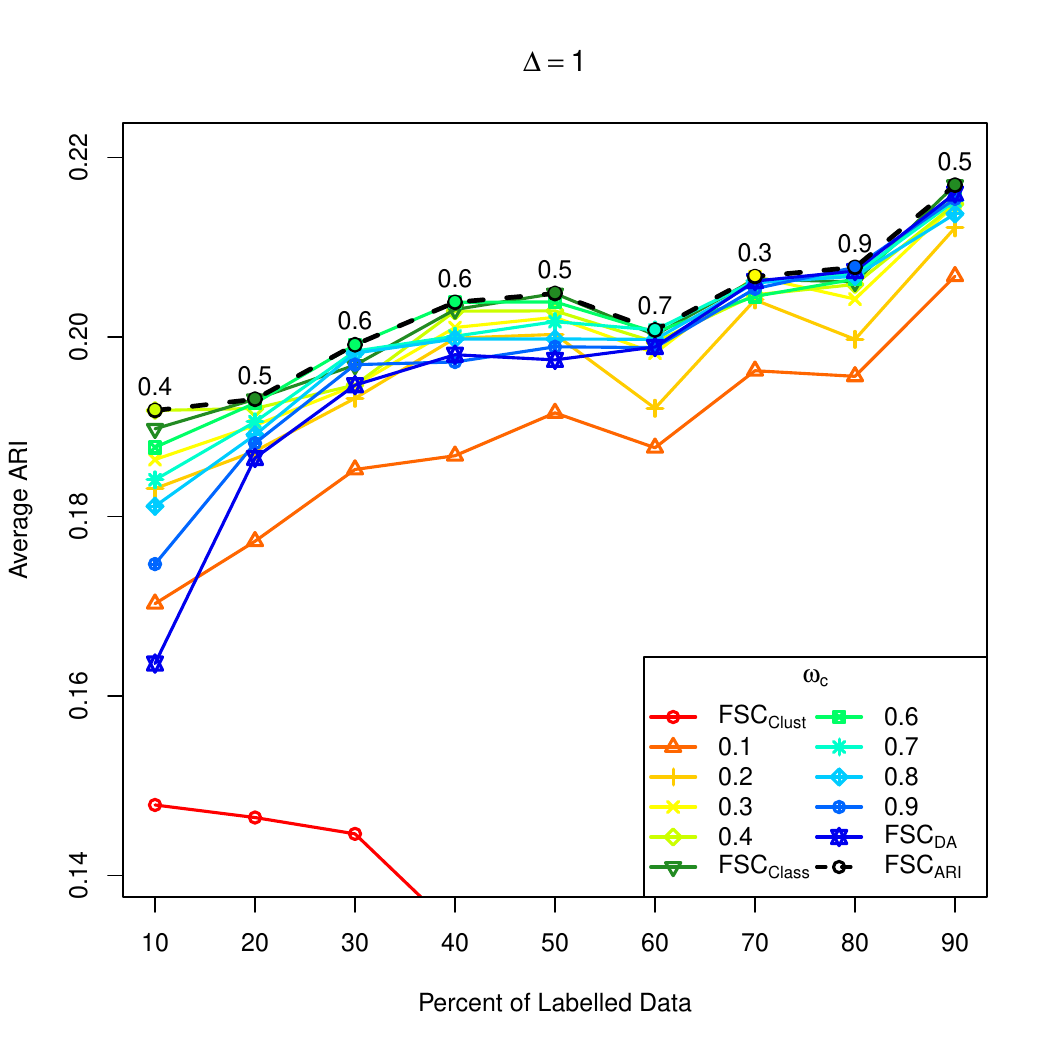} \
\includegraphics[width=0.49\textwidth]{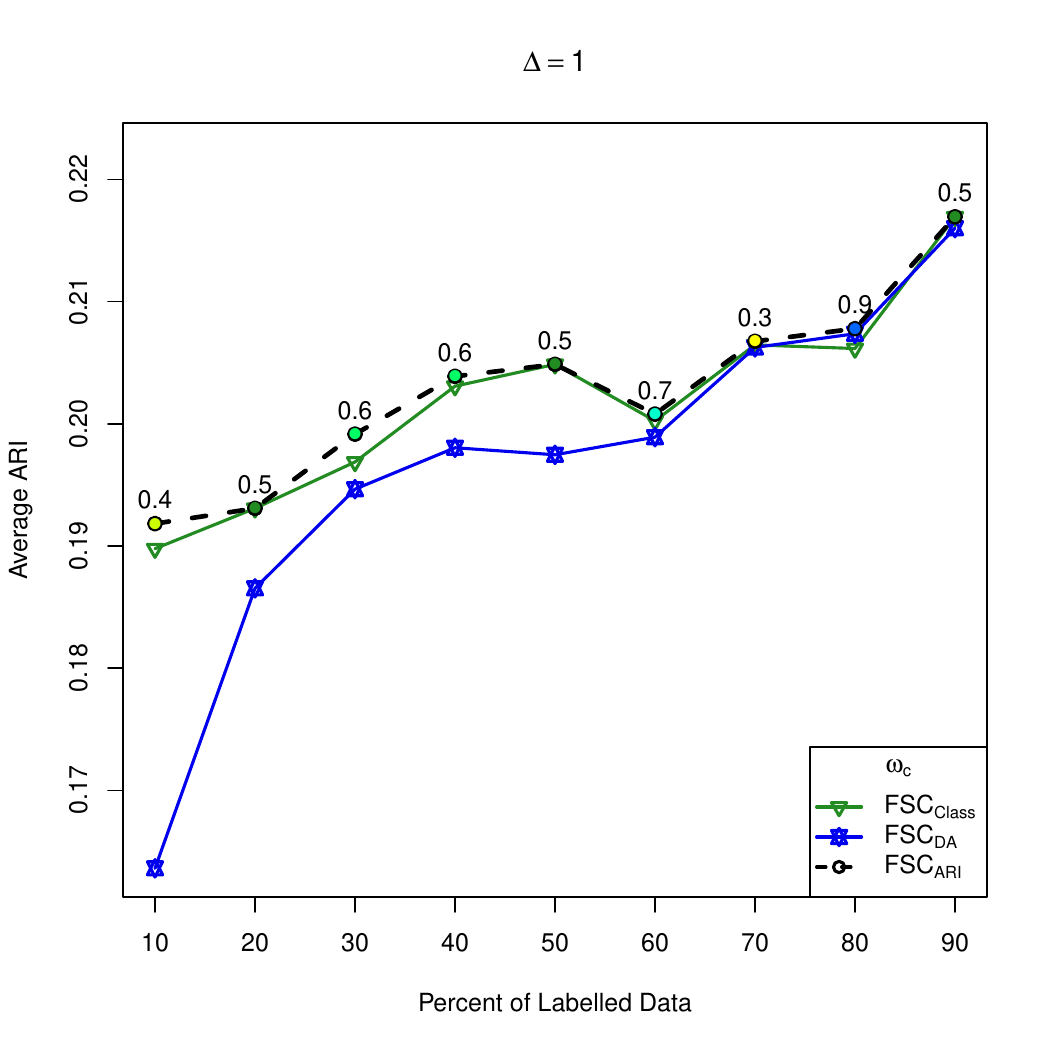}
\end{center}
\vspace{-0.12in}
\caption{
Average ARI values (taken over 100 runs) for FSC$_{\awhat}$  when applied to Simulation 1 with $\Delta = 1$ for $\awhat \in \{\awhatvec, \ariopt\}$ (left) and $\awhat \in \{0.5, 1, \ariopt\}$  (right).}
\label{d5plots}
\end{figure}

\begin{figure}[!h]
\vspace{-0.12in}
\begin{center}
\includegraphics[width=0.49\textwidth]{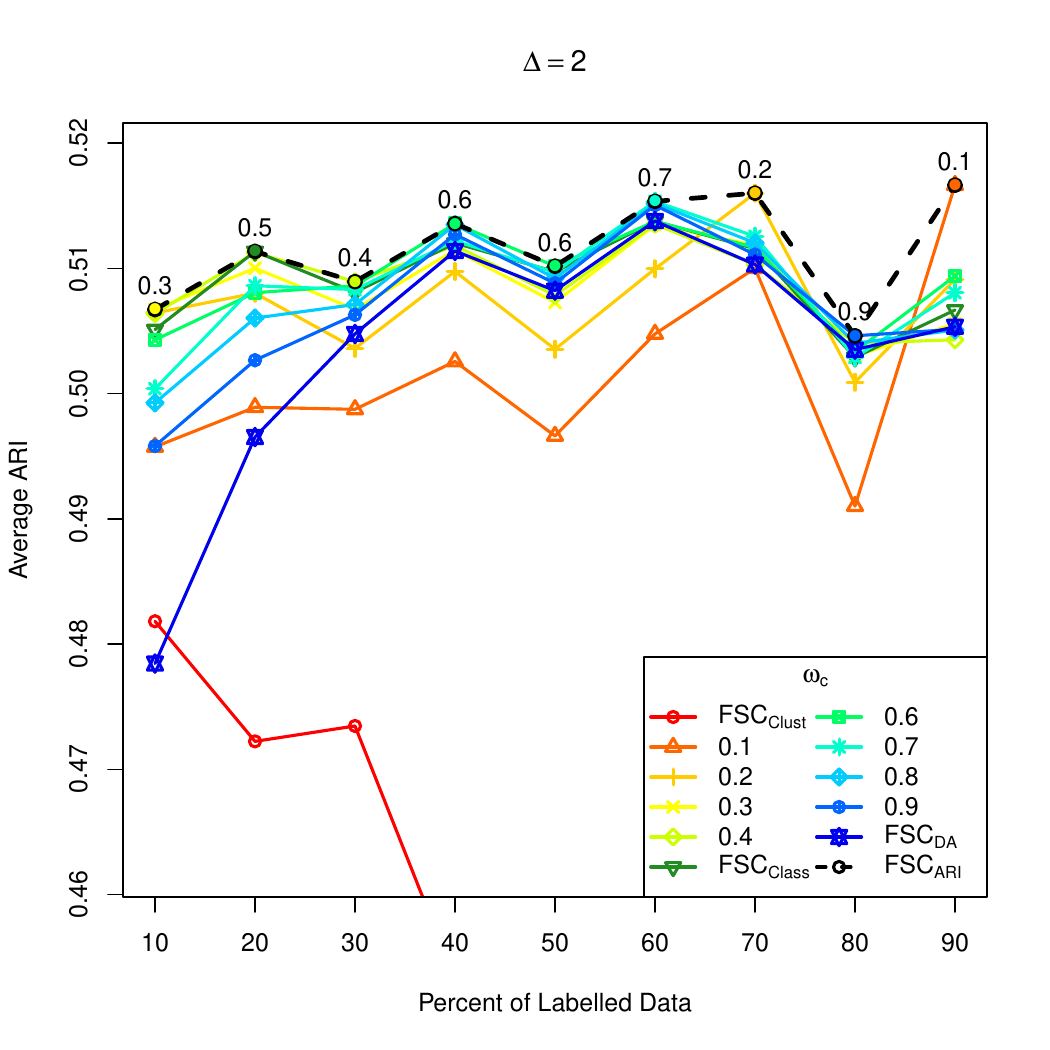} \
\includegraphics[width=0.49\textwidth]{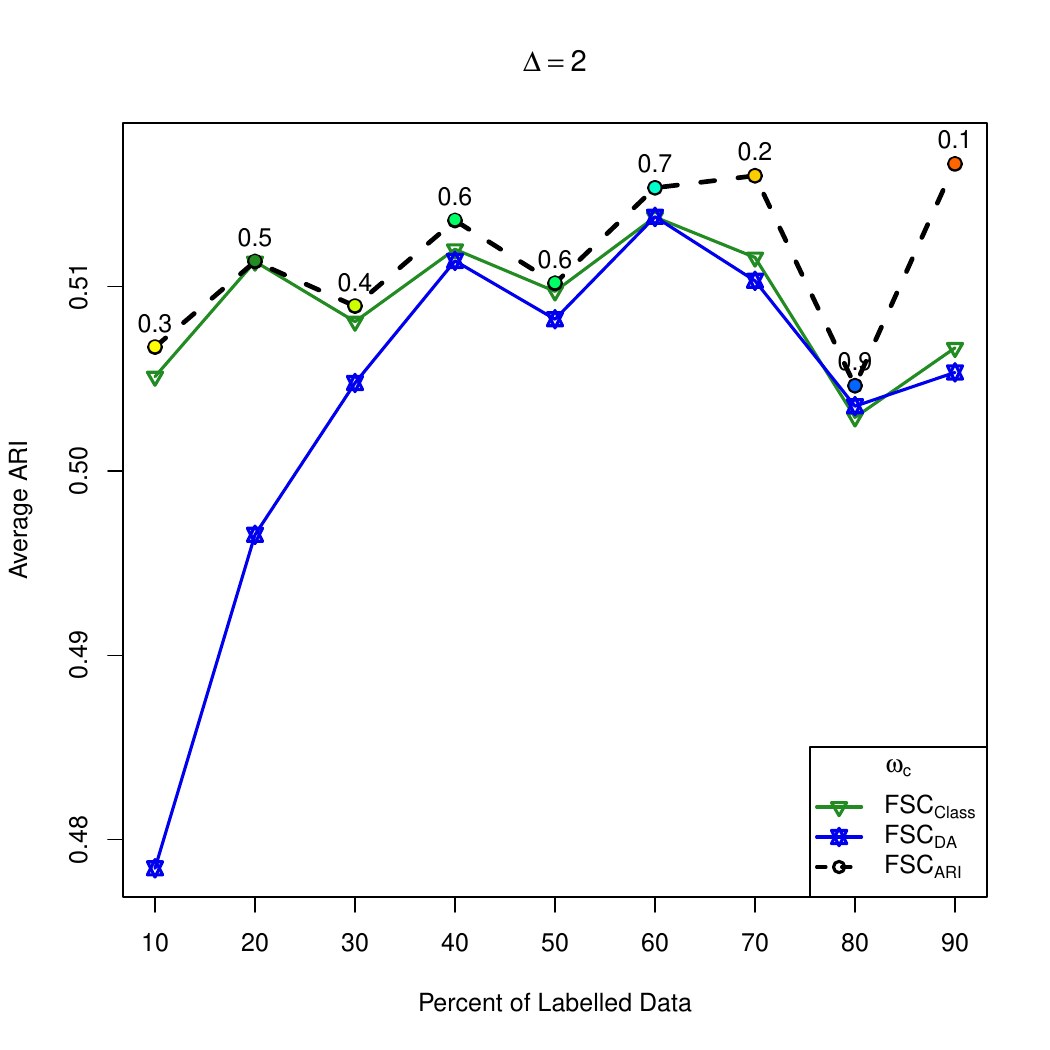}
\end{center}
\vspace{-0.12in}
\caption{
Average ARI values (taken over 100 runs) for FSC$_{\awhat}$  when applied to Simulation 1 with $\Delta = 2$ for $\awhat \in \{\awhatvec, \ariopt\}$ (left) and $\awhat \in \{0.5, 1, \ariopt\}$  (right).}
\label{delta2}
\end{figure}

\begin{figure}[!h]
\vspace{-0.12in}
\begin{center}
\includegraphics[width=0.374\textwidth]{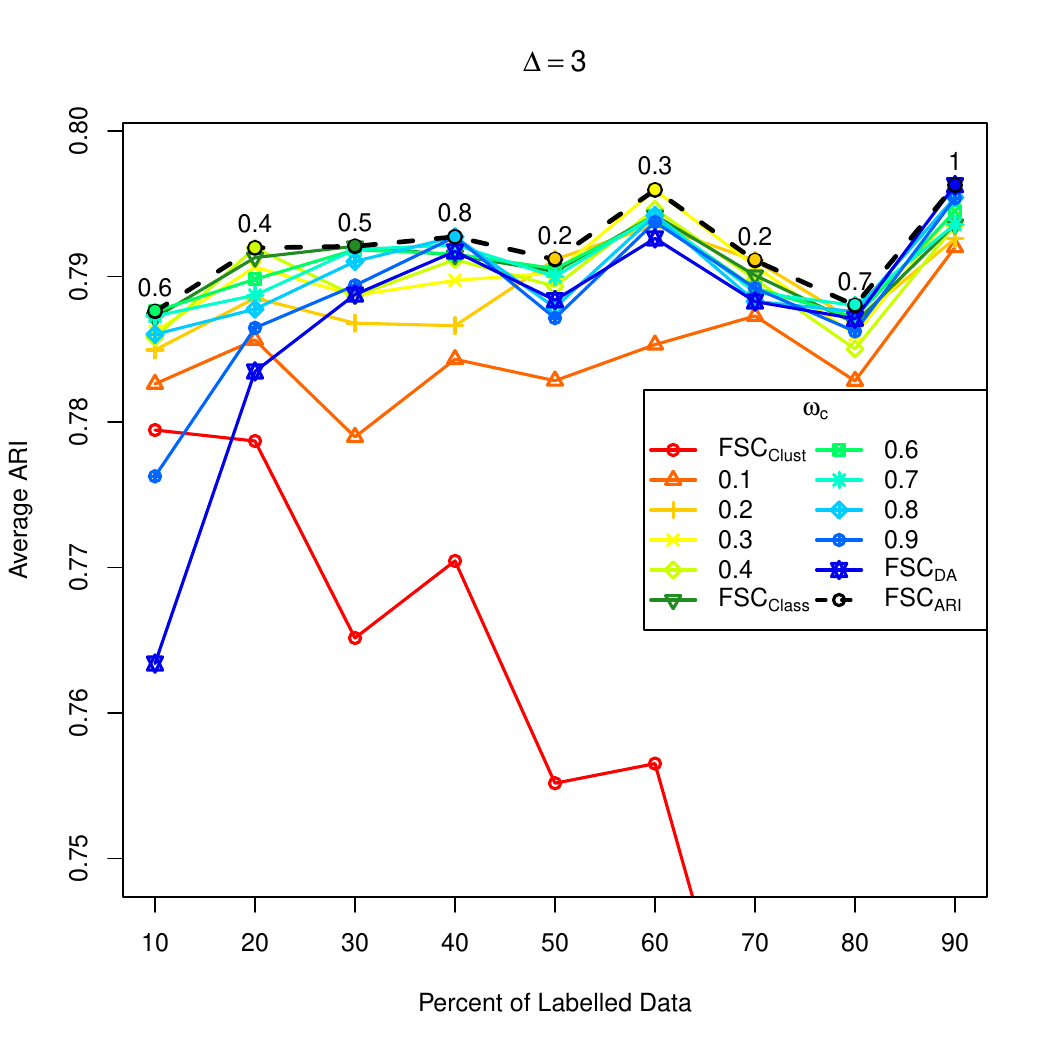} \
\includegraphics[width=0.374\textwidth]{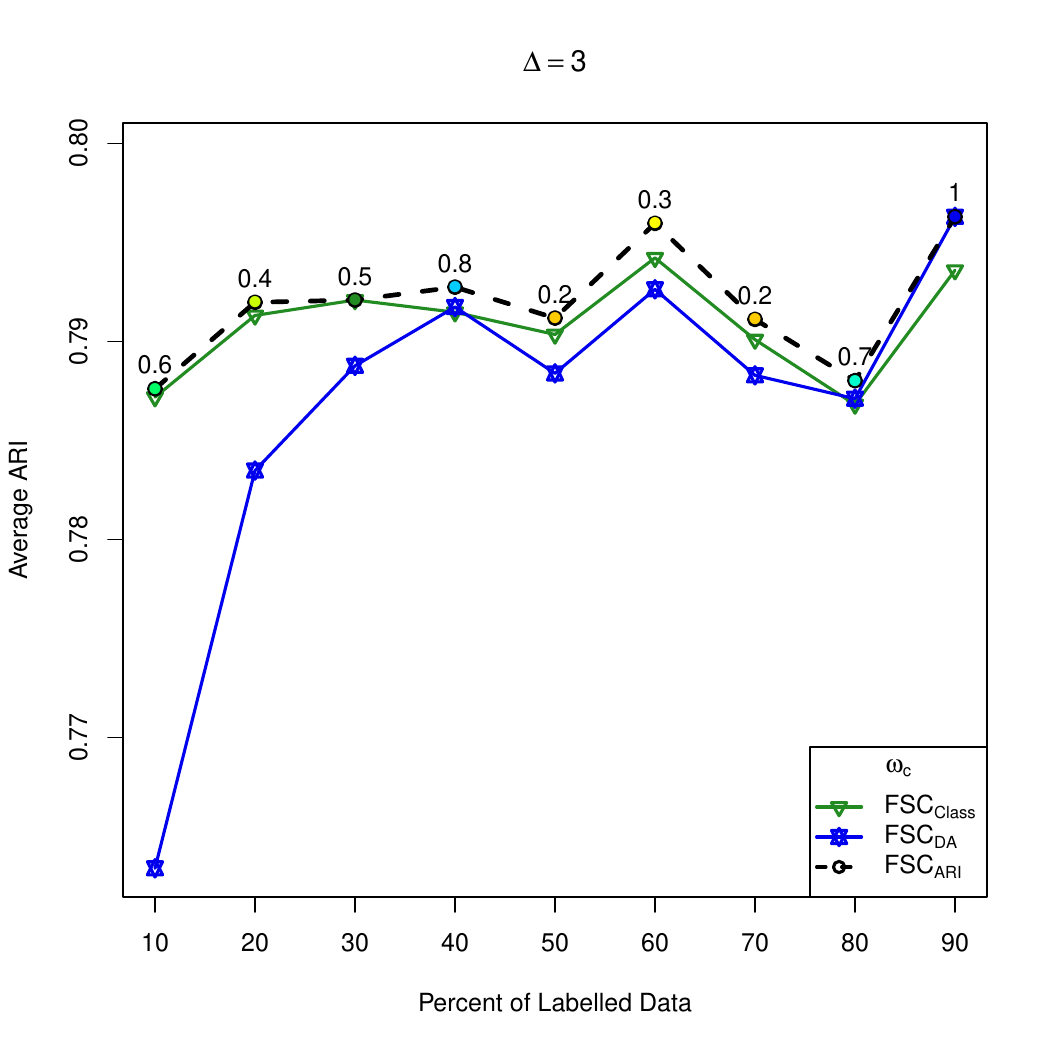}\\
\includegraphics[width=0.374\textwidth]{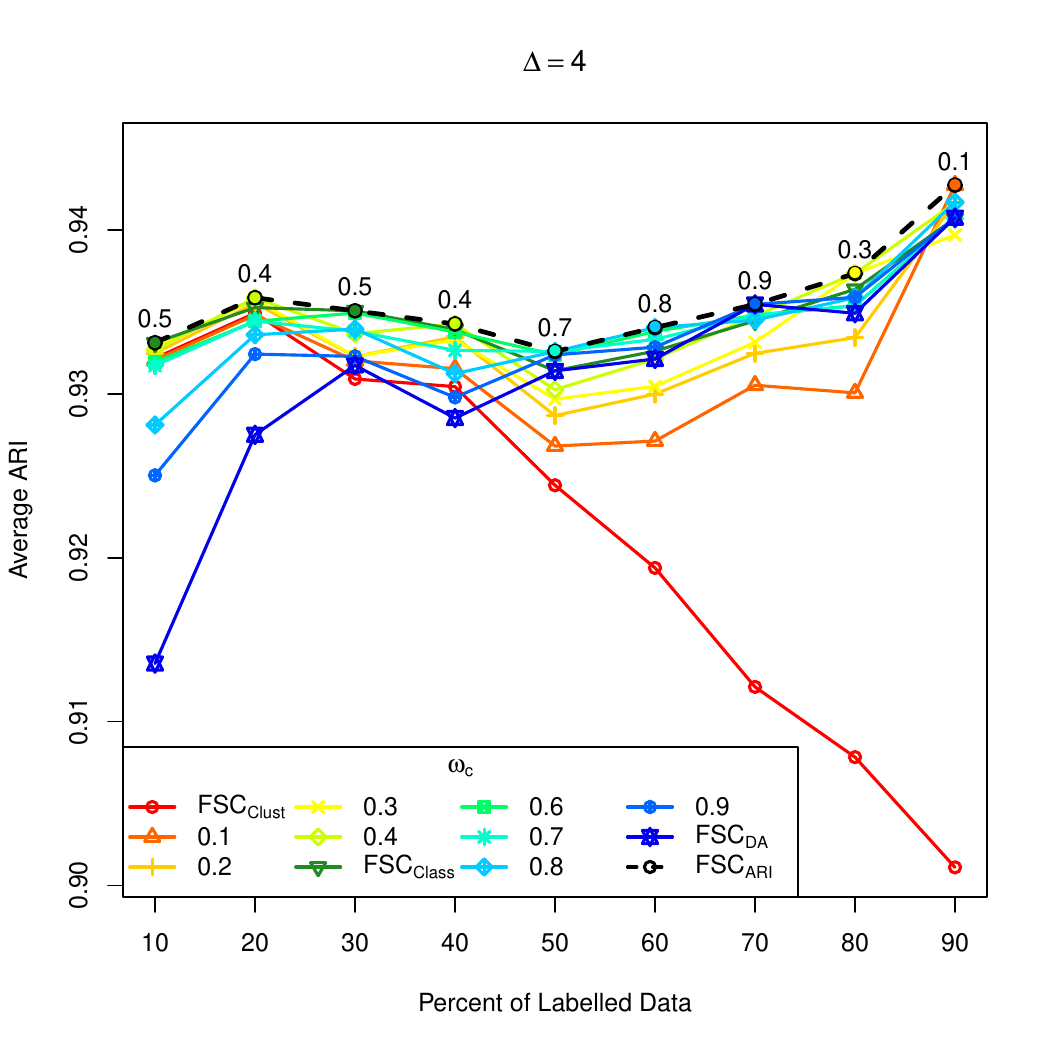} \
\includegraphics[width=0.374\textwidth]{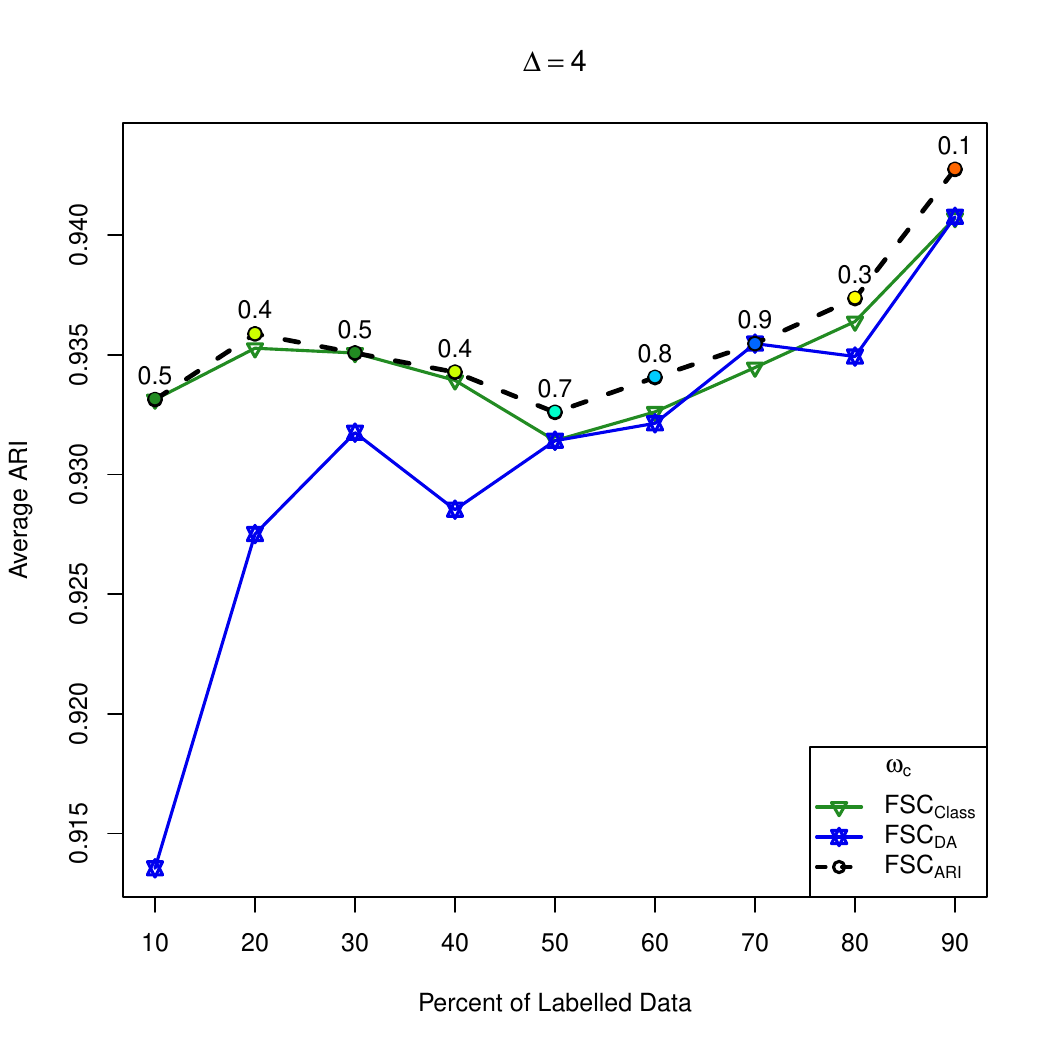}\\
\includegraphics[width=0.374\textwidth]{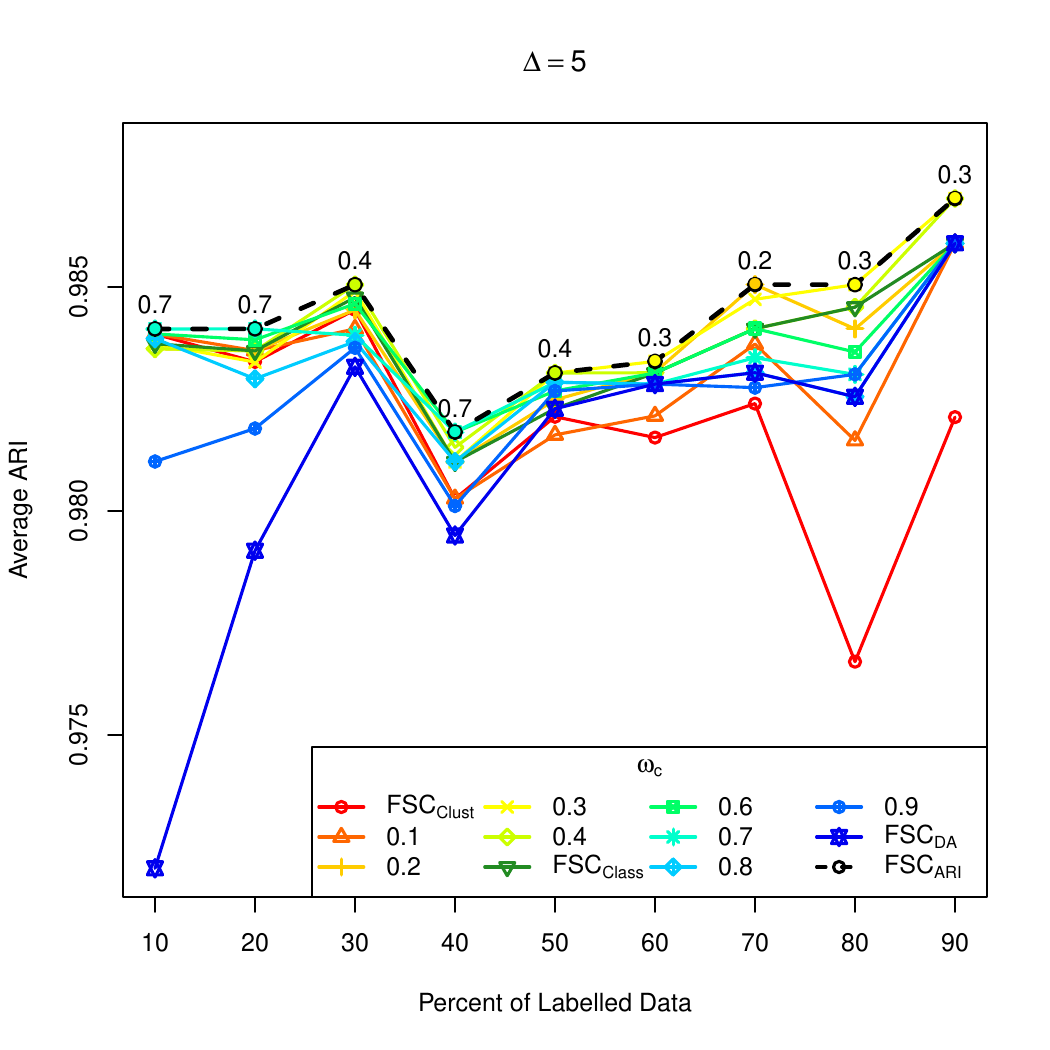} \
\includegraphics[width=0.374\textwidth]{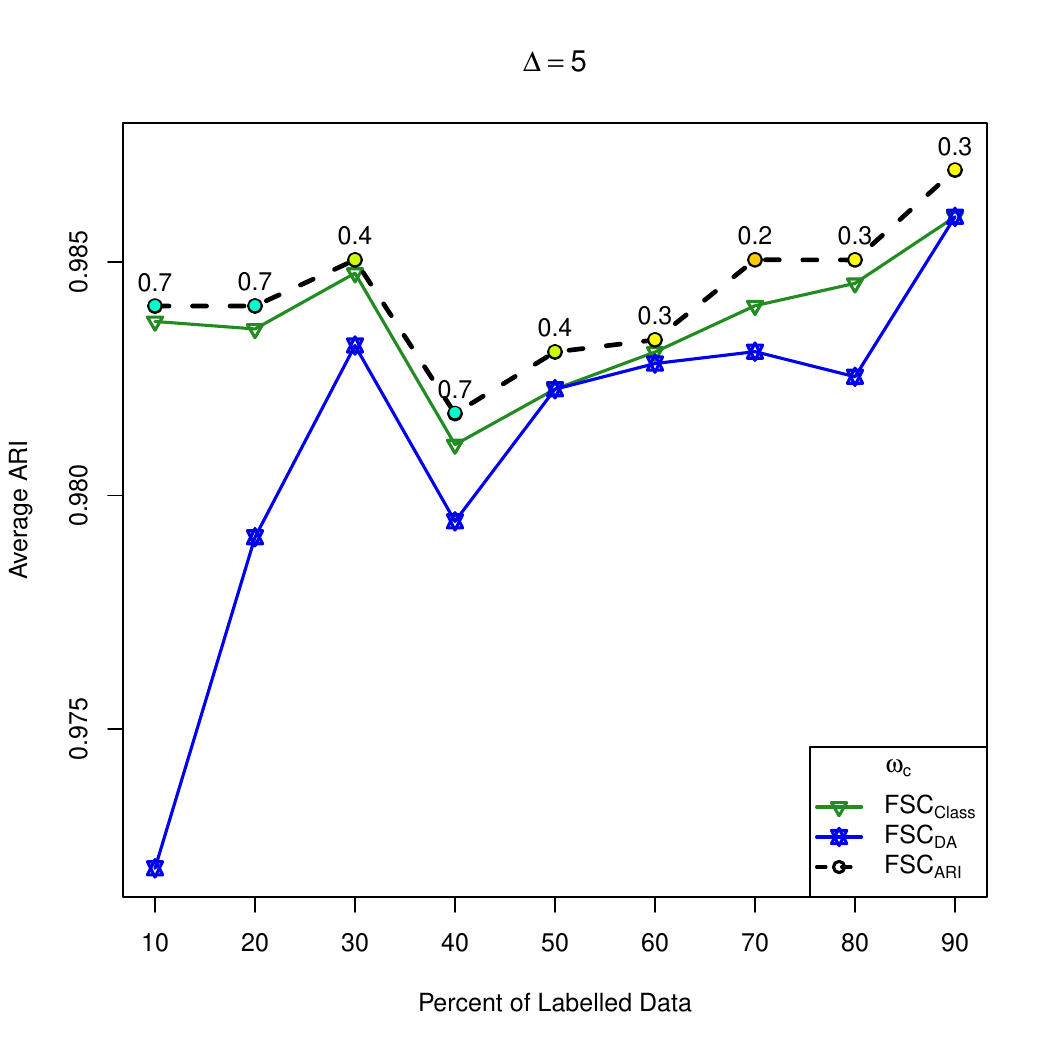}
\end{center}
\vspace{-0.12in}
\caption{
Average ARI values (taken over 100 runs) for FSC$_{\awhat}$  when applied to Simulation 1 with $\Delta = 3,4$ and $5$, respectively, for $\awhat \in \{\awhatvec, \ariopt\}$ (left) and $\awhat \in \{0.5, 1, \ariopt\}$  (right).}
\label{delta345}
\end{figure}

The left-hand side panel of Figure \ref{d5plots} plots $\avgariD$ for all of the fitted FSC models when  $\Delta=1$.  
 The right-hand side panel isolates the main species of interest for ease of viewing.  For scaling purposes, these results may omit model-based clustering (i.e., $\fclust$), which often performed quite poorly.  Included in these plots are the resulting average ARI values for $\fari$; the corresponding $\ariopt$ values are labelled directly on the graph. 
The corresponding plots for $\Delta \in\{2, 3,4,5\}$ are given in Figures \ref{delta2} and \ref{delta345}.

%
%

When applied to the simulation with the highest degree of overlap ($\Delta=1$), 
$\ariopt$ corresponds to model-based classification in three of the nine splits considered. In other words, when 20, 50 and 90\% of the data are labelled, $\fclass$ performs best on average.  In all other cases, there exists a `non-species' value --- i.e., a $\ac \notin \{0,0.5,1\}$ --- that achieves the highest average ARI. For the remaining 36 $\ariopt$ values (nine values of $p$ times four values of $\Delta$), 31 are non-species.  For the remaining five cases, four correspond to model-based classification ($\fclass$) and one to DA ($\fda$). It is worth mentioning that the value of $\ariopt$ is less than 0.5 in 18 of the 31 cases. These cases illustrate how softening the role of labelled observations can actually lead to an increase in classification performance. For instance, the largest gain in ARI is observed at $\Delta=2$ and $p=90$, where $\text{FSC}_{0.1}$ produces an average ARI of 0.517.  In comparison, $\fclust$, $\fclass$, and $\fda$ yield average ARI values of 0.364, 0.507, and 0.505, respectively.
Although the majority of these runs show that non-species tend to produce the highest average ARI scores, we remark that not much difference is observed between $\fclass$ and $\fari$.  
This simulation seems to suggest that when the postulated model is correct,
intermediate weights do not offer a significant improvement over $\fclass$.  
 The next simulation investigates the case in which the model is misspecified. 
\FloatBarrier

\subsection{Simulation 2}
\label{sim2}
\begin{figure}[h]
\vspace{-0.12in}
\begin{center}
\includegraphics[width=0.49\textwidth]{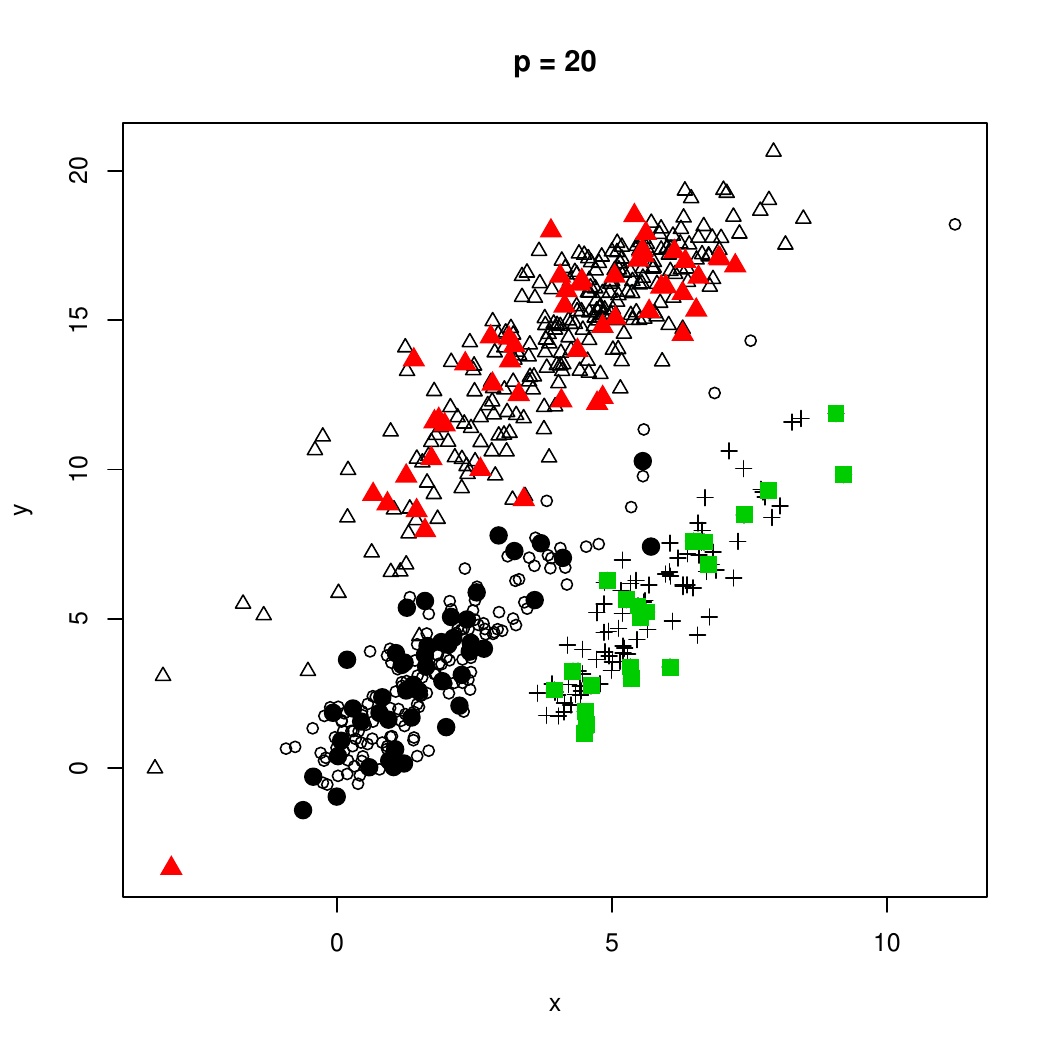} \
\includegraphics[width=0.49\textwidth]{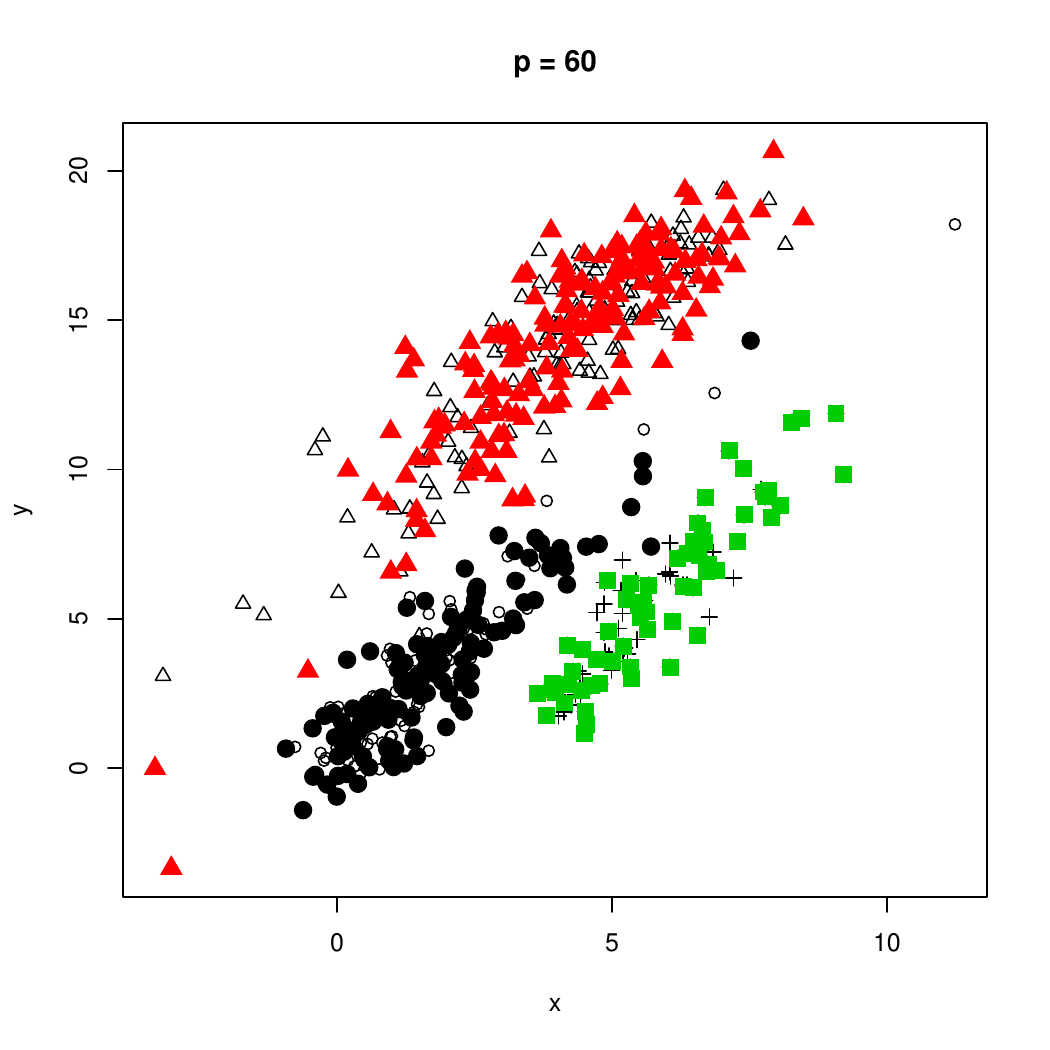}
\end{center}
\vspace{-0.12in}
\caption{Scatter plots for a typical data set simulated according to Simulation 2 with 20\% labelled observations (left) and 60\% labelled observations (right).  Labelled observations are denoted by bold symbols and color represents component membership.}
\label{fig:sim2}
\end{figure}

In this section, we consider 100 simulations  generated from a three-component bivariate `restricted' skew-$t$ mixture \cite[cf.][]{vrbik2012analytic, lee2013}. The component densities are specified to have locations $\bs\xi_1 = (0\quad 0)^\top , \bs\xi_2 = (6\quad 18)^\top , \bs\xi_3 = (4\quad 2)^\top $; skewness $\bs\lambda_1 = (2\quad 4)^\top$, $\bs\lambda_2 = (-2\quad 4)^\top$, $\bs\lambda_3 = (2\quad 4)^\top $; degrees of freedom $\nu_1=10, \nu_2 = 8, \nu_3 = 70$; and scale matrices 
\begin{equation*}
\bs\Sigma  _1= \left( \begin{array}{ccc}
0.4 & 0.2\\
0.2 & 0.5\end{array} \right),\qquad
\bs\Sigma  _2= \left( \begin{array}{ccc}
1 & 0.5\\
0.5 & 1\end{array} \right),\qquad
\bs\Sigma  _3= \left( \begin{array}{ccc}
0.2 & 0\\
0 & 0.3\end{array} \right).
\end{equation*}
Each data set $D_{rp}$ $(r= 1, \dots, 100, p=10,20, \dots, 90)$ contains 600 observations ($n_1 = 200$, $n_2 = 300$, $n_3 =100$).  Typical simulated data sets with 20\% ($p=20$) and 60\% ($p=60$) labelled observations, respectively, are given in Figure \ref{fig:sim2}.

\begin{figure}[h]
\vspace{-0.12in}
\begin{center}
\includegraphics[width=0.49\textwidth]{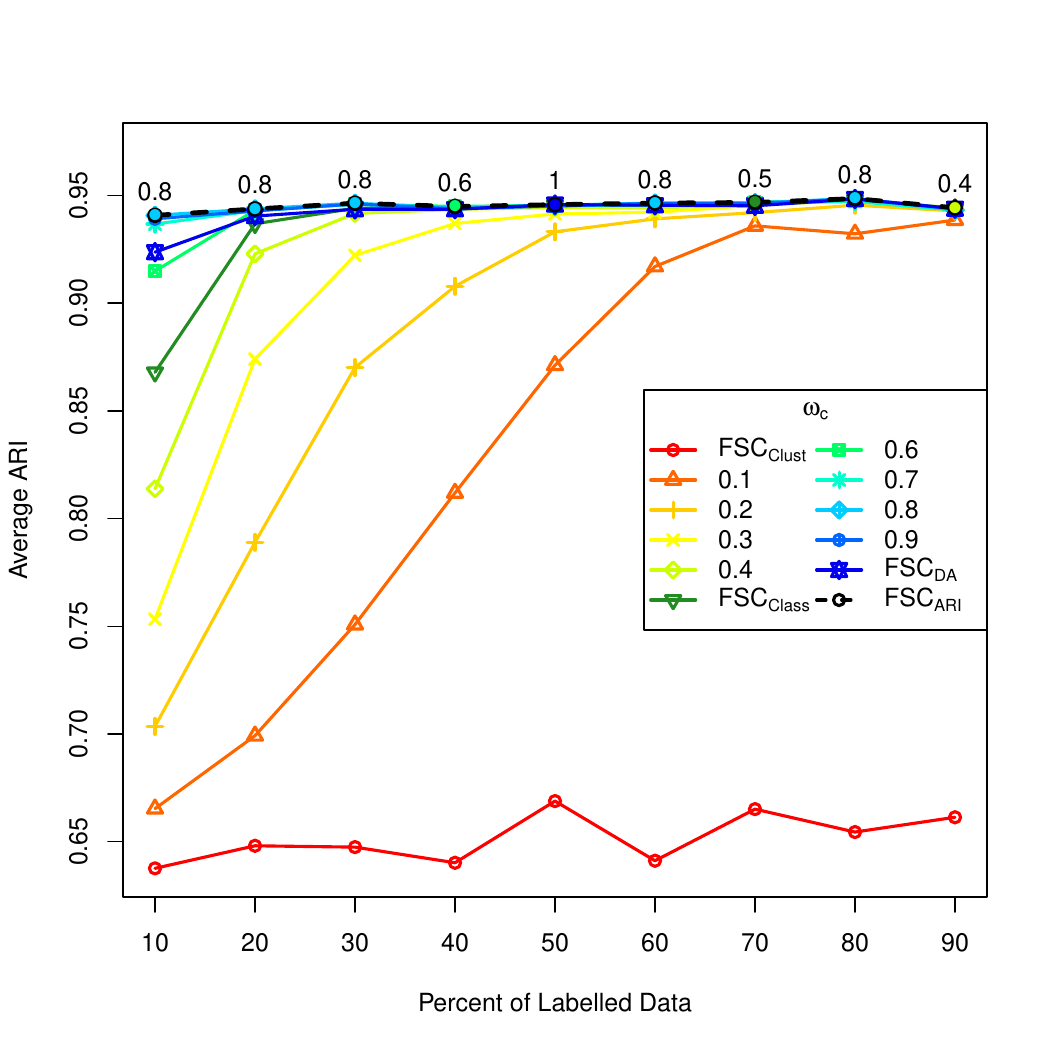} \
\includegraphics[width=0.49\textwidth]{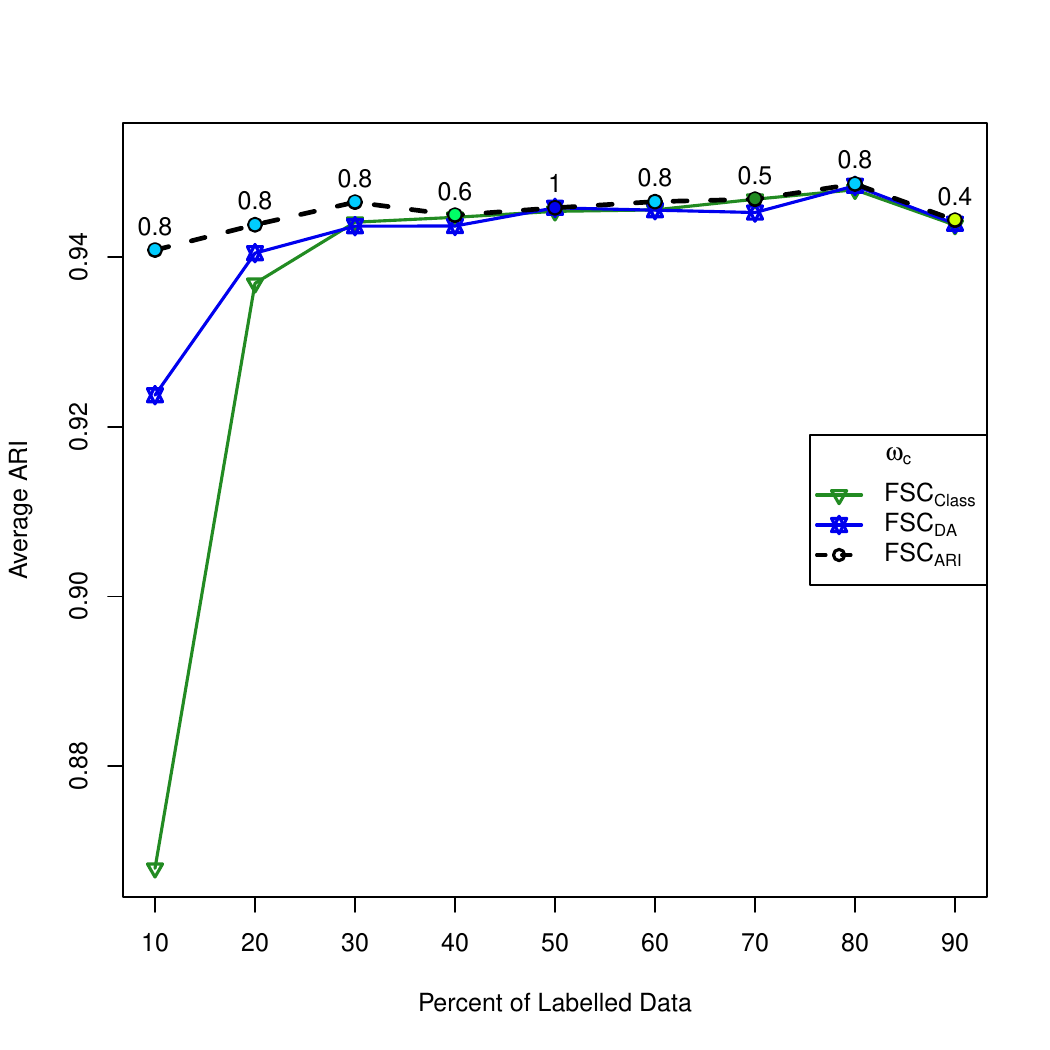}
\end{center}
\vspace{-0.12in}
\caption{
Average ARI values (taken over 100 runs) for FSC$_{\awhat}$  when applied to Simulation 2 for $\awhat \in \{\awhatvec, \ariopt\}$ (left) and $\awhat \in \{0.5, 1, \ariopt\}$  (right).
}
\label{d2plots}
\end{figure}

In this simulation, the average ARI values (Figure \ref{d2plots}) show that FSC performs better for larger values of $\ac$ when $p<70$. After this threshold, there is little difference in performance between the competing methods. As seen in Figure~\ref{d2plots}, when 10\% of the observations are labelled,  $\fclass$ --- which obtains an average ARI score of 0.868 --- performs worse than $\fda$ $(\overline{\text{ARI}}(1, D_{\cdot 10}) = 0.924)$ and $\fari$ ($\overline{\text{ARI}}(\ariopt{}, D_{\cdot 10})=0.941)$. $\fda$  and $\fclass$ are the optimal models when 50 and 70\% of the data, respectively, are labelled.  
As indicated by the plotted values of $\ariopt$, some cases ($p=10,20,30,40,50,80$) benefit from emphasizing the relative role of $D_\text{L}$ over $D_\text{u}$, whereas other cases ($p=90$) perform better when the opposite is true.  The remaining split {(when $p=70$) results in an optimum value of  $\ariopt=0.5$, indicating that $D_\text{L}$ and $D_\text{u}$ are weighted equally when building the classifier}.  

The sporadic patterns observed in these simulations demonstrate the difficulty behind obtaining a general rule for the specification of weights. 
{We continue this investigation in Section~\ref{sec:App} using real data.}

\section{Applications}
\label{sec:App}

In this section, we demonstrate how $\text{FSC}_{\awhat}$ with  $\awhat \in \awhatvec$   compares with the three species of classification when applied to three real data sets. For each data set, 100 random splits are considered; again, we use $p= 10, 20, \dots, 90.$  As in Section \ref{sec:sim1}, the information for each run $r$  with $p\%$ labelled is stored in $D_{rp}$ for $r=1,2,\dots, 100, p = 10, 20, \dots, 90$.
The resulting average ARI values for the competing methods are calculated using \eqref{avgari}.

\subsection{Italian Wine Data}\label{sec:wine}
\begin{figure}[!h]
\vspace{-0.12in}
\begin{center}
\includegraphics[width=0.49\textwidth]{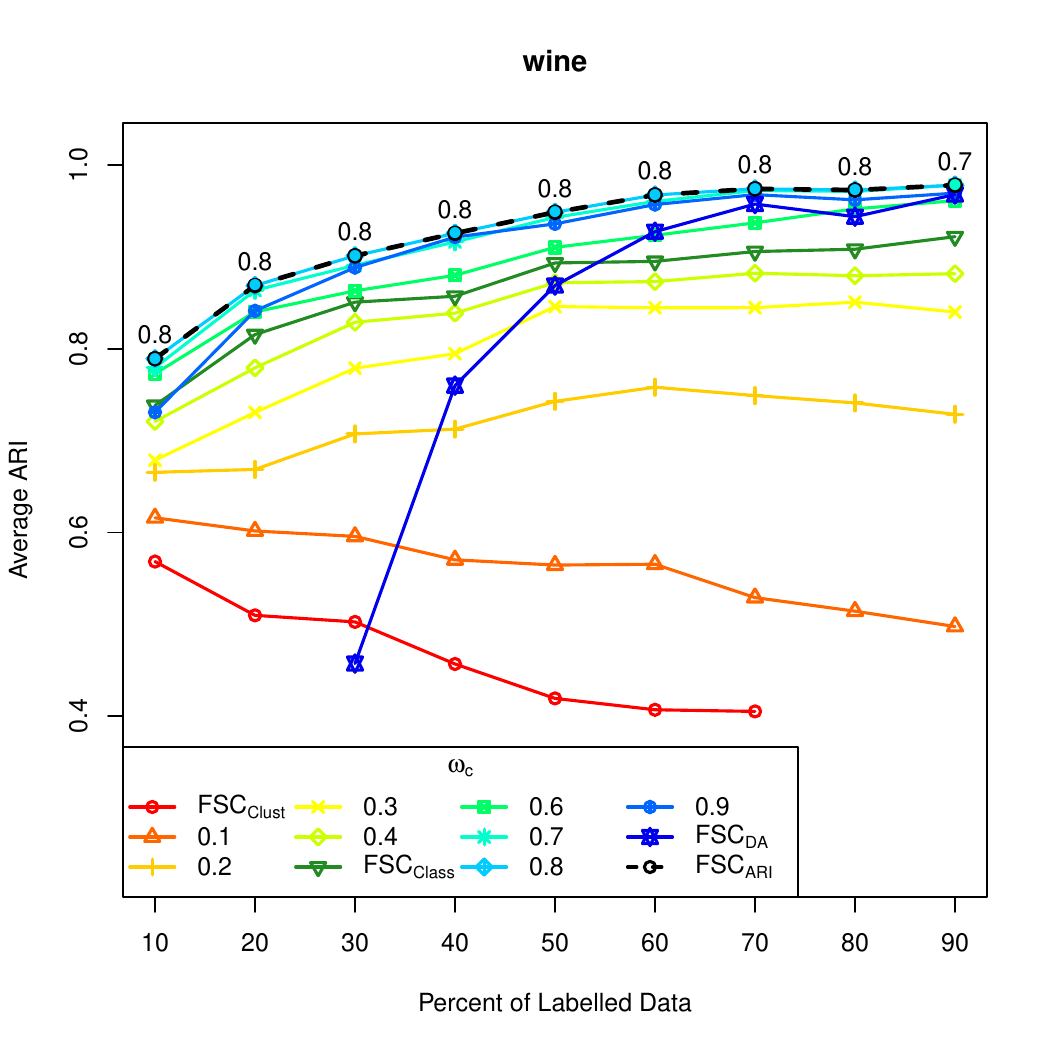} \
\includegraphics[width=0.49\textwidth]{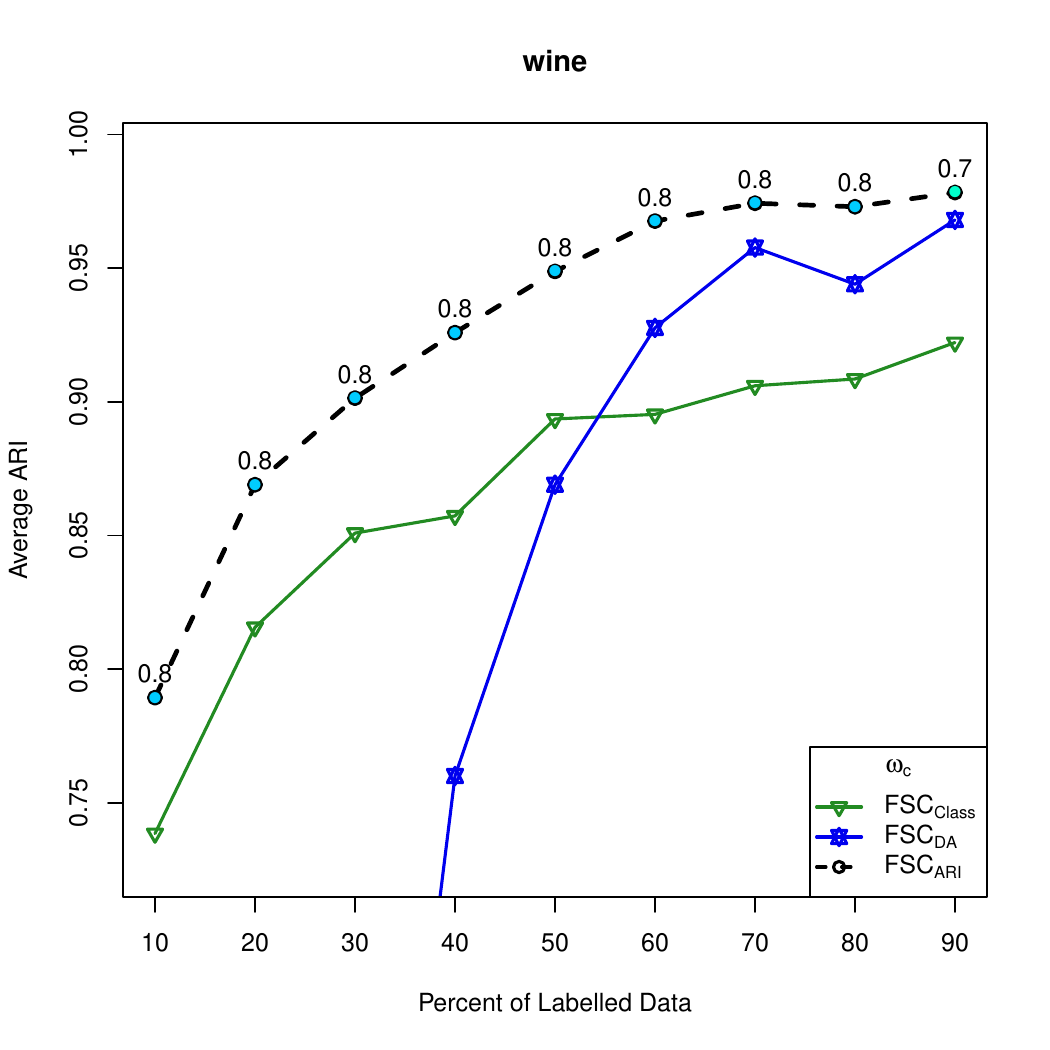} 
\end{center}
\vspace{-0.12in}
\caption{
Average ARI values (taken over 100 runs) for FSC$_{\awhat}$  when applied to the wine data  for $\awhat \in \{\awhatvec, \ariopt\}$ (left) and $\awhat \in \{0.5, 1, \ariopt\}$  (right).}
\label{wine}
\end{figure}

The Italian wine data \citep{forina86} consist of 13 chemical and physical properties of 178 wines from three different grape cultivars. This data set can be accessed through the {\tt gclus} package \citep{gclus} for {\sf R} \citep{Rteam}.  The values of $\avgari$ for the competing methods are plotted in Figure~\ref{wine}. Note that  data are too scarce to run $\fda$ on less than 30\% of the data. Consequently, the support of $\fda$ is the set $\{30, 40, \dots, 90\}$.  Similarly, $\fclust$ is only supported for $p \in \{10, 20, \dots, 70\}$.  

 When $p\leq50$, $\fclass$ performs better than $\fda$; however, when more than half of the data are labelled, $\fda$ outperforms $\fclass$ (Figure~\ref{wine}). The clustering results obtained by $\fari$ are consistently better than for any species of classification, hence all values of $\ariopt$ --- as indicated by the labels in Figure~\ref{wine} --- are non-species.  In fact, for all but one split ($p=90)$, {the optimal weight is 0.8.}
   The largest gain in performance can be seen at $p=40$, where $\fhat{{0.8}}$ attains an average ARI of 0.926.  In comparison, the average ARI values attained by $\fclust$, $\fclass$, and $\fda$ are  0.457, 0.857, and  0.760, respectively.  

\subsection{Crabs Data}\label{sec:crabs}
\begin{figure}[!h]
\vspace{-0.12in}
\begin{center}
\includegraphics[width=0.49\textwidth]{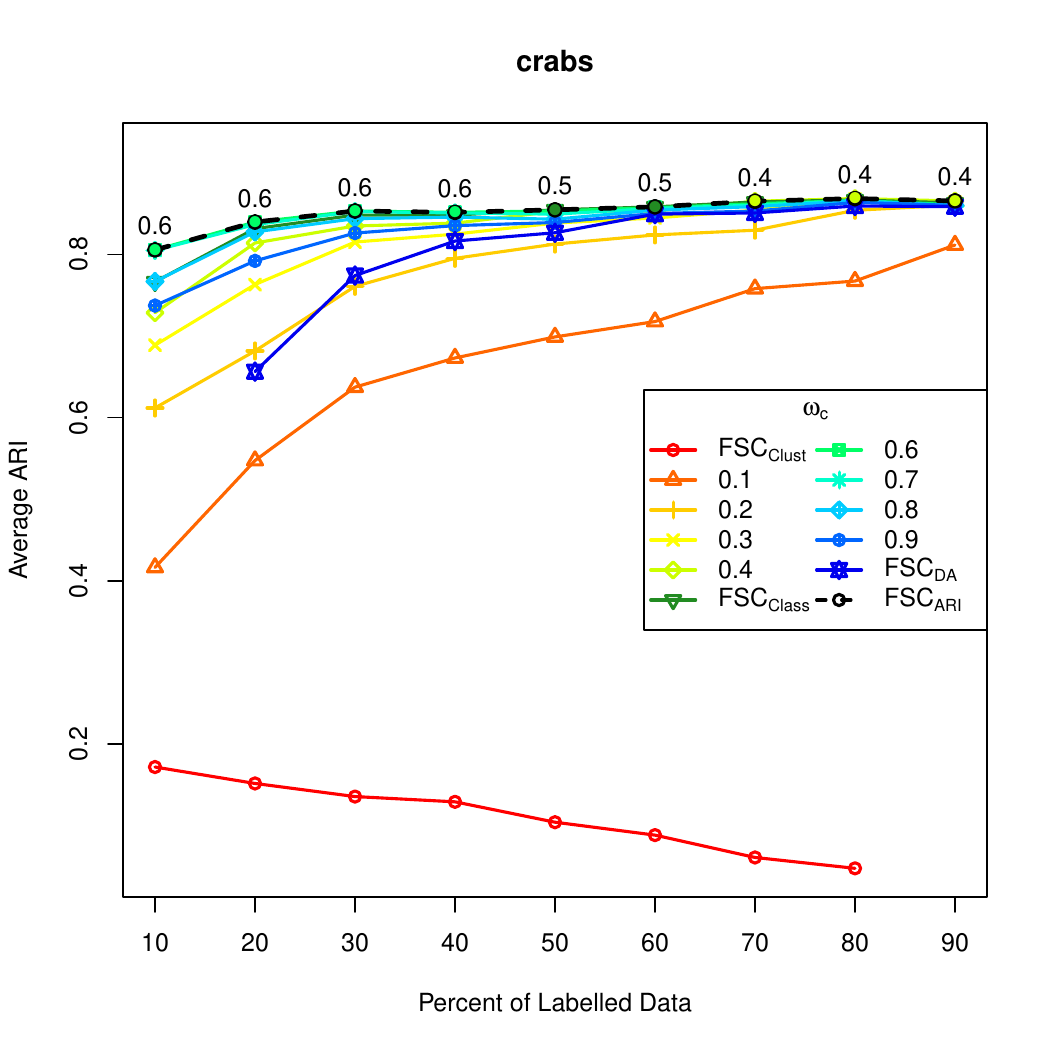} \ 
\includegraphics[width=0.49\textwidth]{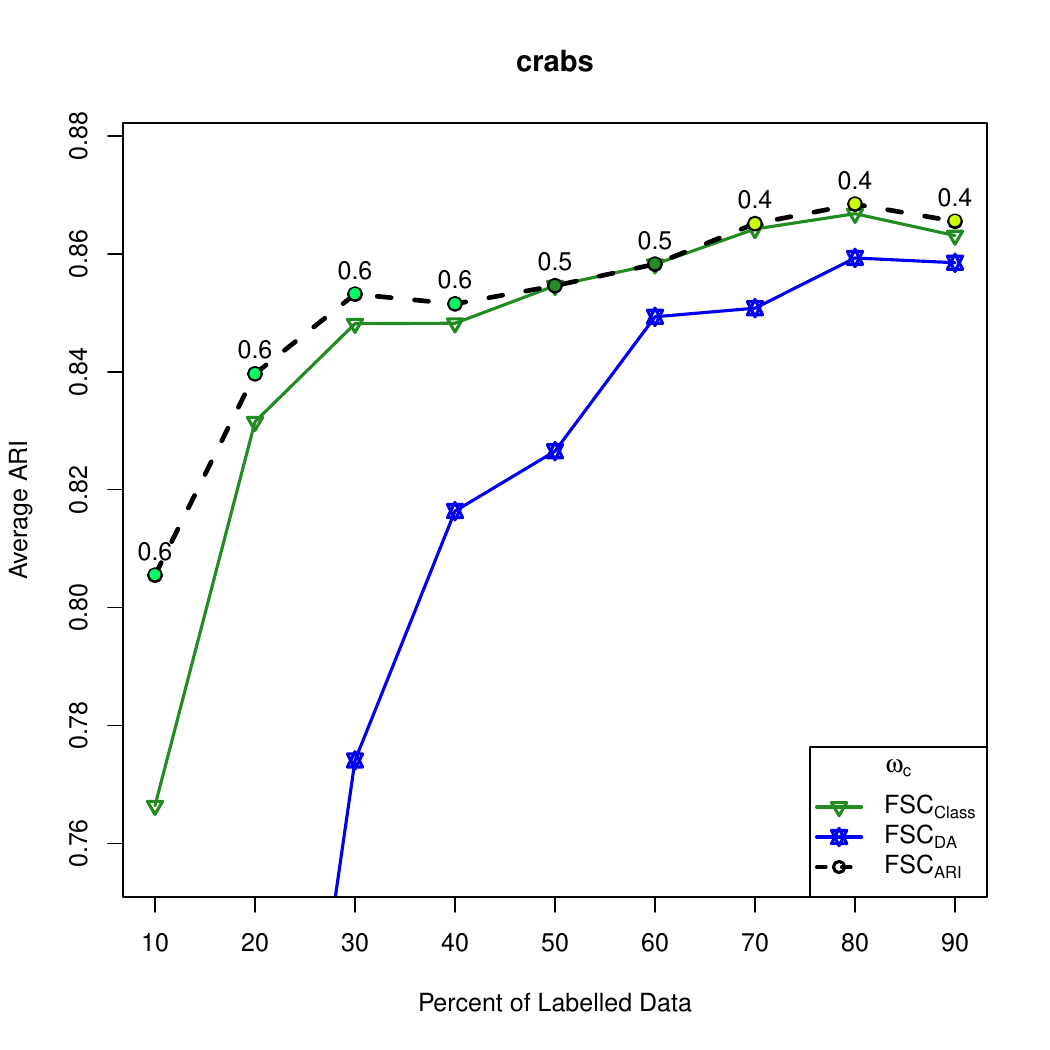}
\end{center}
\vspace{-0.12in}
\caption{
Average ARI values (taken over 100 runs) for FSC$_{\awhat}$  when applied to the crabs data  for $\awhat \in \{\awhatvec, \ariopt\}$ (left) and $\awhat \in \{0.5, 1, \ariopt\}$  (right).}
\label{crabs}
\end{figure}

The crabs data \citep{campbell} contain measurements on the frontal lobe size, rear width, carapace length, carapace width, and body depth of four different types of crab. This data set can be accessed through the {\tt MASS}  package \citep{MASS} for {\sf R}.  
As seen in Figure~\ref{crabs}, model-based classification consistently outperforms DA. Furthermore, $\fclass$ gives comparable performance to $\fari$ for $p \geq 20$.   When $p=10$, slight gains in ARI can be seen by adopting FSC with $\awhat=0.6$. Namely, $\text{FSC}_{0.6}$, $\fclass$, and $\fclust$ obtain { $\avgari$ values of   
0.805, 0.766,   and 0.172, respectively}. Note that there are insufficient data to run $\fda$ with only 10\% labelled observations.
Unlike the previous example, two of the optimal choices for $\ac$ correspond to a particular species of classification; namely, when $p=50, 60$,  $\fari$ is equivalent  to $\fclass$.  For all other values of $p$,  $\ariopt$ is either 0.6 (when $p=10,20,30,40$) or 0.4 (when $p=70,80,90$),  indicating more and less weight, respectively, on labelled observations. 

\subsection{Iris Data}\label{sec:iris}
\begin{figure}[!h]
\vspace{-0.12in}
\begin{center}
\includegraphics[width=0.49\textwidth]{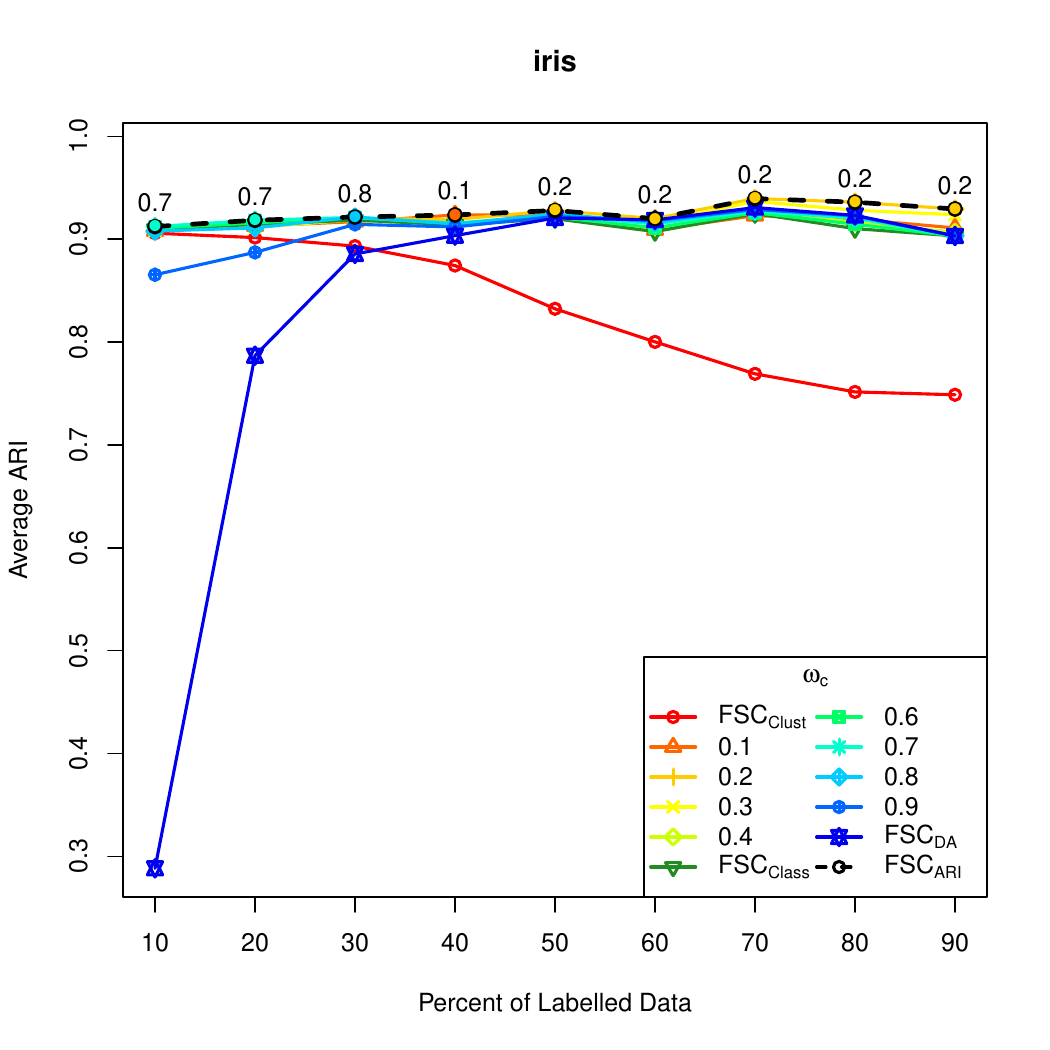} \
\includegraphics[width=0.49\textwidth]{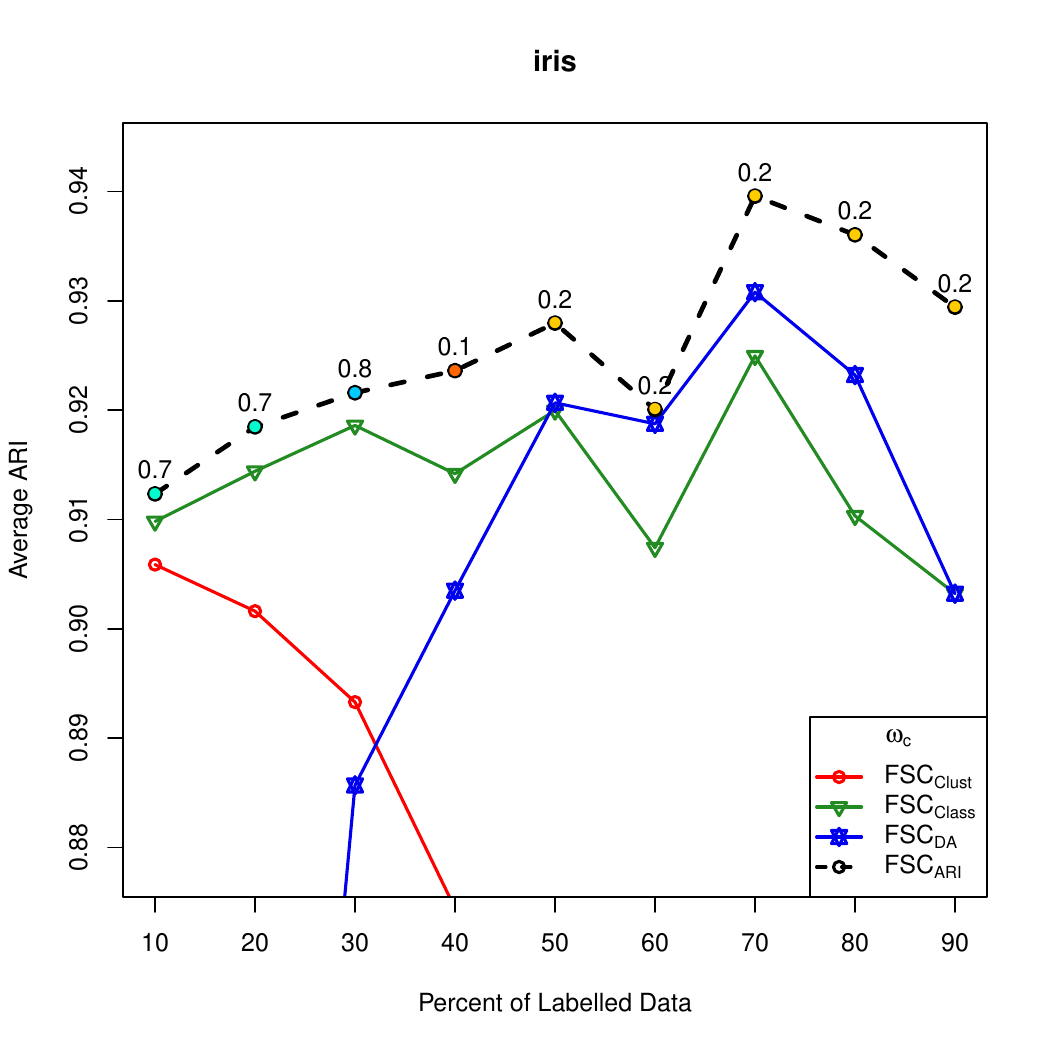} 
\end{center}
\vspace{-0.12in}
\caption{
Average ARI values (taken over 100 runs) for FSC$_{\awhat}$  when applied to the iris data  for $\awhat \in \{\awhatvec, \ariopt\}$ (left) and $\awhat \in \{0, 0.5, 1, \ariopt\}$  (right).}
\label{iris}
\end{figure}

The iris data set contains the length and width, in centimetres, of the sepal and petals of three species of irises \citep{anderson, fisher}.  The iris data are among the base data sets available in {\sf R}.  
The average ARI produced by $\fari$ consistently outperforms all three species (Figure~\ref{iris}).  
For example, $\fhat{0.2}$ yields an average ARI of 0.929 when $p=90$   whereas $\fclust$, $\fclass$, and $\fda$ yield average ARIs of 0.749, 0.903, and 0.903, respectively.  
Note that when $p \leq 30$, all values of $\ariopt$  are  above 0.5, meaning  more weight is given to $D_\text{L}$.  On the contrary, when more than 30\% of observations are labelled, the optimal weights are small ($\ariopt$ = 0.1 or 0.2). This indicates that enhancing the role of  $D_\text{u}$ over $D_\text{L}$ in FSC can lead to an increase in classification performance. 

 \section{Conclusion}\label{sec:con}

Herein, a  flexible classification paradigm, FSC, has been introduced. Constructed by adding weights to the traditional likelihood for semi-supervised classification, FSC allows any level of supervision ranging from unsupervised to supervised, with three levels coinciding with widely known species (unsupervised, semi-supervised, and supervised). We illustrated the FSC approach on real and simulated data, and explored situations where FSC can advantageously be employed.  
Model-based classification (i.e., $\fclass$) proved competitive in some cases; however, for the majority of settings there existed a value of $\ac \notin \{0,0.5,1\}$ for which $\text{FSC}_\ac$ outperformed the three species, namely, supervised ($\fda$), semi-supervised ($\fclass$), and unsupervised classification ($\fclust$).  

Of course, the efficacy of FSC depends on the challenging task of specifying the weight $\ac$. 
The optimal value for $\ac$ --- as determined by the average ARI from a candidate set of potential weights $\awhatvec$ --- has proven to be fairly unpredictable,
reflecting the complicated role that unlabelled and labelled data can have on building a classifier.  
 Specifically, we observed four scenarios: (a) $0<\ariopt < 0.5$; (b) $\ariopt = 0.5$; (c) $0.5 < \ariopt < 1$; and (d) $\ariopt = 1$.  Scenarios (a) and (c) correspond to cases where the best model is obtained by emphasizing the relative roles of unlabelled and labelled data, respectively.  Scenario~(b) corresponds to the optimal model occurring at $\fclass$. Scenario~(d) coincides with $\fda$, and occurs when the inclusion of unlabelled data leads to a decrease in classification performance.  We never observed a scenario in which $\ariopt = 0$ was optimal. 

In conclusion, FSC provides a flexible framework that unites the three species of classification (i.e., supervised, semi-supervised, and unsupervised) and promotes the exploration and development of weighted likelihood techniques. 
 Moving forward, the major challenge is to devise an automated way of selecting $\ac$ that, ideally, never performs worse than the three species.  To this end, the development of a data-driven selection criterion is a subject of ongoing investigation.  
 
 Apart from the delicate issue of weight specification, numerous expansions can be explored within the FSC framework.  For example, one could extend FSC to include the special case where known class labels ($\mz_1$),  rather than labeled observation pairs ($D_\text{L} = \{\mx_1, \mz_1\}$), are ignored.  This might be accomplished by having separate weights for $\mx_i$ and $\mz_i$.
 Another avenue of research involves adapting these models so that they may be applied to situations where observations are systematically unlabelled.  In this way, FSC could help offset the potential bias introduced by systematic missingness.  Expanding on this idea, one might include the possibility of more than two sources of data.  For instance, in addition to unlabelled and labelled data we may allow for a third source, perhaps comprised of outliers or data suspected to be mislabelled.   We believe the methods presented in this paper are simply the forerunners to a different approach to classification, one based upon the weighted likelihood. 

{\small
\subsection*{Acknowledgements}
This work was supported by an Ontario Graduate Scholarship (Vrbik), an Early Researcher Award from the Government of Ontario (McNicholas), and a Discovery Grant from the Natural Sciences and Engineering Research Council of Canada (McNicholas).}


\begin{thebibliography}{}

\bibitem[\protect\citeauthoryear{Akaike}{Akaike}{1973}]{akaike1973information}
Akaike, H. (1973).
\newblock Information theory and an extension of the maximum.
\newblock In {\em Second International Symposium on Information Theory}, pp.\
  267--281.

\bibitem[\protect\citeauthoryear{Anderson}{Anderson}{1935}]{anderson}
Anderson, E. (1935).
\newblock The irises of the {G}asp\'{e} {P}eninsula.
\newblock {\em Bulletin of the American Iris Society\/}~{\em 59}, 2--5.

\bibitem[\protect\citeauthoryear{Andrews, McNicholas, and Subedi}{Andrews
  et~al.}{2011}]{andrews2011}
Andrews, J.~L., P.~D. McNicholas, and S.~Subedi (2011).
\newblock Model-based classification via mixtures of multivariate
  t-distributions.
\newblock {\em Computational Statistics and Data Analysis\/}~{\em 55\/}(1),
  520--529.

\bibitem[\protect\citeauthoryear{Baluja}{Baluja}{1998}]{baluja1998}
Baluja, S. (1998).
\newblock Probabilistic modeling for face orientation discrimination: Learning
  from labeled and unlabeled data.
\newblock In {\em Neural Information Processing Systems (NIPS `98)}, pp.\
  854--860.

\bibitem[\protect\citeauthoryear{Baranchik}{Baranchik}{1970}]{baranchik1970}
Baranchik, A.~J. (1970).
\newblock A family of minimax estimators of the mean of a multivariate normal
  distribution.
\newblock {\em The Annals of Mathematical Statistics\/}~{\em 41}, 642--645.

\bibitem[\protect\citeauthoryear{Baudry, Raftery, Celeux, Lo, and
  Gottardo}{Baudry et~al.}{2010}]{baudry10}
Baudry, J.-P., A.~E. Raftery, G.~Celeux, K.~Lo, and R.~Gottardo (2010).
\newblock Combining mixture components for clustering.
\newblock {\em Journal of Computational and Graphical Statistics\/}~{\em
  19\/}(2), 332--353.

\bibitem[\protect\citeauthoryear{Berry}{Berry}{1994}]{berry1994}
Berry, J.~C. (1994).
\newblock Improving the {J}ames-{S}tein estimator using the {S}tein variance
  estimator.
\newblock {\em Statistics and Probability Letters\/}~{\em 20\/}(3), 241--245.

\bibitem[\protect\citeauthoryear{Biernacki, Celeux, and Govaert}{Biernacki
  et~al.}{2000}]{biernacki2000assessing}
Biernacki, C., G.~Celeux, and G.~Govaert (2000).
\newblock Assessing a mixture model for clustering with the integrated
  completed likelihood.
\newblock {\em IEEE Transactions on Pattern Analysis and Machine
  Intelligence\/}~{\em 22\/}(7), 719--725.

\bibitem[\protect\citeauthoryear{Campbell and Mahon}{Campbell and
  Mahon}{1974}]{campbell}
Campbell, N.~A. and R.~J. Mahon (1974).
\newblock A multivariate study of variation in two species of rock crab of
  genus {L}eptograpsus.
\newblock {\em Australian Journal of Zoology\/}~{\em 22}, 417--425.

\bibitem[\protect\citeauthoryear{Castelli and Cover}{Castelli and
  Cover}{1996}]{castelli1996}
Castelli, V. and T.~M. Cover (1996).
\newblock The relative value of labeled and unlabeled samples in pattern
  recognition with an unknown mixing parameter.
\newblock {\em IEEE Transactions on Information Theory\/}~{\em 42\/}(6),
  2102--2117.

\bibitem[\protect\citeauthoryear{Cozman, Cohen, and Cirelo}{Cozman
  et~al.}{2003}]{cozman2003}
Cozman, F.~G., I.~Cohen, and M.~C. Cirelo (2003).
\newblock Semi-supervised learning of mixture models.
\newblock In {\em International Conference on Machine Learning (ICML 2003)},
  pp.\  99--106.

\bibitem[\protect\citeauthoryear{Dempster, Laird, and Rubin}{Dempster
  et~al.}{1977}]{dempster77}
Dempster, A.~P., N.~M. Laird, and D.~B. Rubin (1977).
\newblock Maximum likelihood from incomplete data via the {EM} algorithm.
\newblock {\em Journal of the Royal Statistical Society: Series~B\/}~{\em
  39\/}(1), 1--38.

\bibitem[\protect\citeauthoryear{Edwards and Cavalli-Sforza}{Edwards and
  Cavalli-Sforza}{1965}]{edwards65}
Edwards, A. W.~F. and L.~L. Cavalli-Sforza (1965).
\newblock A method for cluster analysis.
\newblock {\em Biometrics\/}~{\em 21}, 362--375.

\bibitem[\protect\citeauthoryear{Fisher}{Fisher}{1936}]{fisher}
Fisher, R.~A. (1936).
\newblock The use of multiple measurements in taxonomic problems.
\newblock {\em Annals of Eugenics\/}~{\em 7\/}(Part II), 179--188.

\bibitem[\protect\citeauthoryear{Forina, Armanino, Castino, and
  Ubigli}{Forina et~al.}{1986}]{forina86}
Forina, M., C.~Armanino, M.~Castino, and M.~Ubigli (1986).
\newblock Multivariate data analysis as a discriminating method of the origin of wines.
\newblock {\em Vitis\/}~{\em 25}, 189--201.

\bibitem[\protect\citeauthoryear{Hartigan and Wong}{Hartigan and
  Wong}{1979}]{hartigan1979}
Hartigan, J.~A. and M.~A. Wong (1979).
\newblock Algorithm {AS} 136: A k-means clustering algorithm.
\newblock {\em Journal of the Royal Statistical Society: Series~C\/}~{\em
  28\/}(1), 100--108.

\bibitem[\protect\citeauthoryear{Hennig}{Hennig}{2010}]{hennig10}
Hennig, C. (2010).
\newblock Methods for merging {G}aussian mixture components.
\newblock {\em Advances in Data Analysis and Classification\/}~{\em 4}, 3--34.

\bibitem[\protect\citeauthoryear{Hu}{Hu}{1994}]{hu1994}
Hu, F. (1994).
\newblock {\em Relevance Weighted Smoothing and a New Bootstrap Method}.
\newblock Ph.\ D. thesis, University of British Columbia.

\bibitem[\protect\citeauthoryear{Hu}{Hu}{1997}]{hu1997}
Hu, F. (1997).
\newblock The asymptotic properties of the maximum-relevance weighted
  likelihood estimators.
\newblock {\em The Canadian Journal of Statistics\/}~{\em 25}, 45--59.

\bibitem[\protect\citeauthoryear{Hu and Zidek}{Hu and Zidek}{2001}]{hu2001}
Hu, F. and J.~V. Zidek (2001).
\newblock The relevance weighted likelihood with applications.
\newblock In {\em Empirical {B}ayes and Likelihood Inference}, pp.\  211--235.
  Springer.

\bibitem[\protect\citeauthoryear{Hu and Zidek}{Hu and
  Zidek}{2002}]{huZidek2002}
Hu, F. and J.~V. Zidek (2002).
\newblock The weighted likelihood.
\newblock {\em The Canadian Journal of Statistics\/}~{\em 30\/}(3), 347--371.

\bibitem[\protect\citeauthoryear{Hubert and Arabie}{Hubert and
  Arabie}{1985}]{hubert85}
Hubert, L. and P.~Arabie (1985).
\newblock Comparing partitions.
\newblock {\em Journal of Classification\/}~{\em 2}, 193--218.

\bibitem[\protect\citeauthoryear{Hurley}{Hurley}{2012}]{gclus}
Hurley, C. (2012).
\newblock {\em gclus: Clustering Graphics}.
\newblock R package version 1.3.1.

\bibitem[\protect\citeauthoryear{James and Stein}{James and
  Stein}{1961}]{james1961}
James, W. and C.~Stein (1961).
\newblock Estimation with quadratic loss.
\newblock In {\em Proceedings of the Fourth Berkeley Symposium on Mathematical
  Statistics and Probability}, Volume~1, pp.\  361--379.

\bibitem[\protect\citeauthoryear{Kawakita and Takeuchi}{Kawakita and
  Takeuchi}{2014}]{kawakita2014}
Kawakita, M. and J.~Takeuchi (2014).
\newblock Safe semi-supervised learning based on weighted likelihood.
\newblock {\em Neural Networks\/}~{\em 53}, 146--164.

\bibitem[\protect\citeauthoryear{Kullback and Leibler}{Kullback and
  Leibler}{1951}]{kullback1951}
Kullback, S. and R.~A. Leibler (1951).
\newblock On information and sufficiency.
\newblock {\em The Annals of Mathematical Statistics\/}~{\em 22\/}(1), 79--86.

\bibitem[\protect\citeauthoryear{Lee and McLachlan}{Lee and
  McLachlan}{2013}]{lee2013}
Lee, S. and G.~McLachlan (2013).
\newblock On mixtures of skew normal and skew t-distributions.
\newblock {\em Advances in Data Analysis and Classification\/}~{\em 7\/}(3),
  241--266.

\bibitem[\protect\citeauthoryear{McCallum and Nigam}{McCallum and
  Nigam}{1998}]{mccallum1998}
McCallum, A. and K.~Nigam (1998).
\newblock Employing {EM} and pool-based active learning for text
  classification.
\newblock In {\em Proceedings of the 15th International Conference on Machine
  Learning (ICML 1998)}, Madison, pp.\  350--358.

\bibitem[\protect\citeauthoryear{McNicholas}{McNicholas}{2010}]{mcnicholas2010}
McNicholas, P.~D. (2010).
\newblock Model-based classification using latent {G}aussian mixture models.
\newblock {\em Journal of Statistical Planning and Inference\/}~{\em 140\/}(5),
  1175--1181.

\bibitem[\protect\citeauthoryear{Nigam, McCallum, Thrun, and Mitchell}{Nigam
  et~al.}{2000}]{nigam2000}
Nigam, K., A.~K. McCallum, S.~Thrun, and T.~Mitchell (2000).
\newblock Text classification from labeled and unlabeled documents using {EM}.
\newblock {\em Machine Learning\/}~{\em 39\/}(2--3), 103--134.

\bibitem[\protect\citeauthoryear{Plant\'{e}}{Plant\'{e}}{2008}]{plante2008}
Plant\'{e}, J.~F. (2008).
\newblock {\em Adaptive Likelihood Weights and Mixtures of Empirical
  Distributions}.
\newblock Ph.\ D. thesis, The University of British Columbia.

\bibitem[\protect\citeauthoryear{Plant\'{e}}{Plant\'{e}}{2009}]{plante2009}
Plant\'{e}, J.~F. (2009).
\newblock Asymptotic properties of the {MAMSE} adaptive likelihood weights.
\newblock {\em Journal of Statistical Planning and Inference\/}~{\em 139\/}(7),
  2147--2161.

\bibitem[\protect\citeauthoryear{{R Core Team}}{{R Core Team}}{2015}]{Rteam}
{R Core Team} (2015).
\newblock {\em R: A Language and Environment for Statistical Computing}.
\newblock Vienna, Austria: R Foundation for Statistical Computing.

\bibitem[\protect\citeauthoryear{Ratsaby and Venkatesh}{Ratsaby and
  Venkatesh}{1995}]{ratsaby1995}
Ratsaby, J. and S.~S. Venkatesh (1995).
\newblock Learning from a mixture of labeled and unlabeled examples with
  parametric side information.
\newblock In {\em Proceedings of the Eighth Annual Conference on Computational
  Learning Theory}, pp.\  412--417. ACM.

\bibitem[\protect\citeauthoryear{Schwarz}{Schwarz}{1978}]{schwarz1978estimating}
Schwarz, G. (1978).
\newblock Estimating the dimension of a model.
\newblock {\em The Annals of Statistics\/}~{\em 6\/}(2), 461--464.

\bibitem[\protect\citeauthoryear{Scott and Symons}{Scott and
  Symons}{1971}]{scott71}
Scott, A.~J. and M.~J. Symons (1971).
\newblock Clustering methods based on likelihood ratio criteria.
\newblock {\em Biometrics\/}~{\em 27}, 387--397.

\bibitem[\protect\citeauthoryear{Sokolovska, Capp{\'e}, and Yvon}{Sokolovska
  et~al.}{2008}]{sokolovska2008}
Sokolovska, N., O.~Capp{\'e}, and F.~Yvon (2008).
\newblock The asymptotics of semi-supervised learning in discriminative
  probabilistic models.
\newblock In {\em Proceedings of the 25th {I}nternational {C}onference on
  {M}achine {L}earning}, pp.\  984--991. ACM.

\bibitem[\protect\citeauthoryear{Stein}{Stein}{1956}]{stein1956}
Stein, C. (1956).
\newblock Inadmissibility of the usual estimator for the mean of a multivariate
  normal distribution.
\newblock In {\em Proceedings of the Third Berkeley Symposium on Mathematical
  Statistics and Probability}, Volume~1, pp.\  197--206.

\bibitem[\protect\citeauthoryear{Steinley}{Steinley}{2004}]{steinley2004properties}
Steinley, D. (2004).
\newblock Properties of the {H}ubert-{A}rabie adjusted {R}and index.
\newblock {\em Psychological {M}ethods\/}~{\em 9\/}(3), 386.

\bibitem[\protect\citeauthoryear{Strawderman}{Strawderman}{1973}]{strawderman1973}
Strawderman, W.~E. (1973).
\newblock Proper {B}ayes minimax estimators of the multivariate normal mean
  vector for the case of common unknown variances.
\newblock {\em The Annals of Statistics\/}~{\em 6}, 1189--1194.

\bibitem[\protect\citeauthoryear{Vandewalle, Biernacki, Celeux, and
  Govaert}{Vandewalle et~al.}{2008}]{vandewalle2008unlabeled}
Vandewalle, V., C.~Biernacki, G.~Celeux, and G.~Govaert (2008).
\newblock Are unlabeled data useful in semi-supervised model-based
  classification? {C}ombining hypothesis testing and model choice.
\newblock In {\em Proceedings of the first joint meeting of the Soci{\'e}t{\'e}
  Francophone de Classification and the Classification and Data Analysis Group
  of SIS}, pp.\  433--436.

\bibitem[\protect\citeauthoryear{Venables and Ripley}{Venables and
  Ripley}{2002}]{MASS}
Venables, W.~N. and B.~D. Ripley (2002).
\newblock {\em Modern Applied Statistics with S\/} (Fourth ed.).
\newblock New York: Springer.

\bibitem[\protect\citeauthoryear{Vrbik and McNicholas}{Vrbik and
  McNicholas}{2012}]{vrbik2012analytic}
Vrbik, I. and P.~D. McNicholas (2012).
\newblock Analytic calculations for the {EM} algorithm for multivariate skew-t
  mixture models.
\newblock {\em Statistics and Probability Letters\/}~{\em 82\/}(6), 1169--1174.

\bibitem[\protect\citeauthoryear{Vrbik and McNicholas}{Vrbik and
  McNicholas}{2014}]{vrbik14}
Vrbik, I. and P.~D. McNicholas (2014).
\newblock Parsimonious skew mixture models for model-based clustering and
  classification.
\newblock {\em Computational Statistics and Data Analysis\/}~{\em 71},
  196--210.

\bibitem[\protect\citeauthoryear{Wald}{Wald}{1949}]{wald}
Wald, A. (1949).
\newblock Note on the consistency of the maximum likelihood estimate.
\newblock {\em The Annals of Mathematical Statistics\/}~{\em 20\/}(4),
  595--601.

\bibitem[\protect\citeauthoryear{Wang}{Wang}{2001}]{wang2001}
Wang, S.~X. (2001).
\newblock {\em Maximum Weighted Likelihood Estimation}.
\newblock Ph.\ D. thesis, University of British Columbia.

\bibitem[\protect\citeauthoryear{Wang, van Eeden, and Zidek}{Wang
  et~al.}{2004}]{wang2004}
Wang, S.~X., C.~van Eeden, and J.~V. Zidek (2004).
\newblock Asymptotic properties of maximum weighted likelihood estimators.
\newblock {\em Journal of Statistical Planning and Inference\/}~{\em 119},
  37--54.

\bibitem[\protect\citeauthoryear{Wang and Zidek}{Wang and
  Zidek}{2005}]{wang2005}
Wang, S.~X. and J.~V. Zidek (2005).
\newblock Selecting likelihood weights by cross-validation.
\newblock {\em Annals of Statistics\/}~{\em 33}, 463--501.

\bibitem[\protect\citeauthoryear{Wang}{Wang}{2006}]{wang2006}
Wang, X. (2006).
\newblock Approximating {B}ayesian inference by weighted likelihood.
\newblock {\em Canadian Journal of Statistics\/}~{\em 34\/}(2), 279--298.

\bibitem[\protect\citeauthoryear{Wolfe}{Wolfe}{1965}]{wolfe1965computer}
Wolfe, J.~H. (1965).
\newblock A computer program for the maximum likelihood analysis of types.
\newblock Technical Bulletin 65-15, U.S.\ Naval Personnel Research Activity.

\end{thebibliography}

\end{document}